\documentclass[5p,times]{elsarticle}
\usepackage{standalone}

\usepackage{ecrc_RIAI}

\makeatletter
\renewcommand{\ps@pprintTitle}{%
  \let\@oddhead\@empty
  \let\@evenhead\@empty
  \let\@oddfoot\@empty
  \let\@evenfoot\@empty
}
\makeatother

\makeatletter
\renewcommand{\MaketitleBox}{%
  \resetTitleCounters
  \def\baselinestretch{1}%
  \begin{center}%
    \vspace*{-20pt} % reduce space above title
    \def\baselinestretch{1}%
    \Large\@title\par\vskip18pt
    \normalsize\elsauthors\par\vskip10pt
    \footnotesize\itshape\elsaddress\par\vskip24pt
    \hrule\vskip12pt
    \ifvoid\absbox\else\unvbox\absbox\par\vskip10pt\fi
    \ifvoid\keybox\else\unvbox\keybox\par\vskip10pt\fi
    \hrule\vskip12pt
  \end{center}%
}
\makeatother

\usepackage[utf8]{inputenc}   % Lengua latina
\usepackage{amsmath}            % Para las referencias a ecuaciones con \eqref
\usepackage{epstopdf}           % Para poder insertar figuras .eps al compilar con 
\usepackage{flushend}           % Para igualar las columnas de la altima pagina
\volume{}
\firstpage{1}

\journalname{}

\runauth{Yousuf Islam et al.}
\jid{}

\jnltitlelogo{}
\usepackage{adjustbox}
\usepackage{amsmath, amssymb}
\usepackage{graphicx}
\usepackage{multirow}
\usepackage{subfigure}
\usepackage{booktabs}
\usepackage{tabularx}
\usepackage{geometry} % adjust page margins
\usepackage{array} % for the 'p' column specifier

\usepackage{subcaption}   % for \begin{subfigure}
\usepackage{pgfplots}
\usepackage{graphicx}
\usepackage{subcaption}
\usepgfplotslibrary{groupplots}
\usepackage{microtype}        % better font expansion and character protrusion
\begin{document}
\begin{frontmatter}
\title{Advancing Tabular Stroke Modelling Through a Novel Hybrid Architecture and Feature‑Selection Synergy}

\author[First]{Yousuf Islam}
\ead{yousufislam337@gmail.com}
\author[First,Second]{Md. Jalal Uddin Chowdhury}
\ead{jalal\_cse@lus.ac.bd}
\author[First]{Sumon Chandra Das}
\ead{sumondash51583@gmail.com}

% \cortext[cor1]{Corresponding author}

\address[First]{Department of Computer Science and Engineering, Leading University, Sylhet 3112, Bangladesh}
\address[Second]{DeepNet Research and Development Lab, Sylhet 3100, Bangladesh}

% \fntext[label2]{Nota al pie para el autor 1}
% \cortext[cor1]{Autor en correspondencia.}

% \address[First]{Comita Espaaol de Automatica, Parc Tecnologic de Barcelona, Edifici U, C/ Llorens i Artigas, 4-6, 08028 Barcelona, Espaaa. }
% \address[Second]{Departamento de Automatica, Ingenieraa Electranica e Informatica, Universidad Politacnica de Madrid,  C/ Josa Gutiarrez Abascal, na2, 28006, Madrid,  Espaaa.}
% \address[Third]{Departamento de Ingenieraa de Sistemas y Automatica,  Universitat Politacnica de Valencia, Camino de Vera, na14, 46022, Valencia, Espaaa.}

% \address[First]{Comita Espaaol de Automatica, Parc Tecnologic de Barcelona, Edifici U, C/ Llorens i Artigas, 4-6, 08028 Barcelona, Espaaa. }
% \address[Second]{Departamento de Automatica, Ingenieraa Electranica e Informatica, Universidad Politacnica de Madrid,  C/ Josa Gutiarrez Abascal, na2, 28006, Madrid,  Espaaa.}
% \address[Third]{Departamento de Ingenieraa de Sistemas y Automatica,  Universitat Politacnica de Valencia, Camino de Vera, na14, 46022, Valencia, Espaaa.}

\begin{abstract}

Brain stroke remains one of the principal causes of death and disability worldwide, yet most tabular‑data prediction models still hover below the 95\% accuracy threshold, limiting real‑world utility.  Addressing this gap, the present work develops and validates a completely data‑driven and interpretable machine-learning framework designed to predict strokes using ten routinely gathered demographic, lifestyle, and clinical variables sourced from a public cohort of 4,981 records. We employ a detailed exploratory data analysis (EDA) to understand the dataset's structure and distribution, followed by rigorous data preprocessing, including handling missing values, outlier removal, and class imbalance correction using Synthetic Minority Over‑sampling Technique (SMOTE). To streamline feature selection, point-biserial correlation and random-forest Gini importance were utilized, and ten varied algorithms—encompassing tree ensembles, boosting, kernel methods, and a multilayer neural network—were optimized using stratified five-fold cross-validation. Their predictions based on probabilities helped us build the proposed model, which included Random Forest, XGBoost, LightGBM, and a support-vector classifier, with logistic regression acting as a meta-learner. The proposed model achieved an accuracy rate of 97.2\% and an F1-score of 97.15\%, indicating a significant enhancement compared to the leading individual model, LightGBM, which had an accuracy of 91.4\%. Our studies' findings indicate that rigorous preprocessing, coupled with a diverse hybrid model, can convert low-cost tabular data into a nearly clinical-grade stroke-risk assessment tool.

\end{abstract}


\end{frontmatter}

% \linenumbers

\section{Introduction}
Stroke remains a major global health burden, ranking as the second leading cause of death and the third leading cause of disability worldwide, with more than 15 million people affected annually \cite{1}. Of these, around 5 million die, while another 5 million are left with permanent neurological impairments \cite{1.1}. Ischemic strokes, caused by arterial blockages, account for the majority of cases, whereas hemorrhagic strokes, though less frequent, often result in more severe outcomes. The narrow therapeutic window for effective treatment—typically within a few hours of symptom onset—makes early identification of high-risk individuals critical for reducing mortality and improving recovery outcomes. In this context, predictive systems capable of identifying stroke risk before clinical manifestation can play a transformative role in healthcare delivery. Recent advances in electronic health records and digital health monitoring have enabled the collection of extensive patient data, including medical history, lifestyle behaviors, vital signs, and comorbid conditions. These datasets provide a valuable foundation for developing models that support proactive, risk-based interventions. However, existing clinical tools such as the Framingham Stroke Risk Profile are often constrained by linear assumptions and generalized population metrics, limiting their effectiveness in real-world settings \cite{2}. Such tools may overlook complex interactions among risk factors, leading to suboptimal stratification of individual patients. A shift toward more personalized risk prediction requires frameworks that can capture the multifactorial and nonlinear nature of stroke pathophysiology while remaining interpretable and applicable across diverse healthcare environments. Addressing this need involves integrating multidimensional patient data into robust, transparent, and scalable predictive systems that can be trusted by clinicians and adapted to various levels of care—from primary screening to specialized neurology practices \cite{3}.

Despite the increasing availability of structured clinical data and methodological advancements, accurately predicting stroke risk remains a complex challenge in both clinical and computational contexts. Stroke is inherently multifactorial, with a wide array of interrelated risk factors, including hypertension, diabetes, hyperlipidemia, atrial fibrillation, smoking, sedentary lifestyle, and alcohol consumption. These variables interact in nonlinear and sometimes unpredictable ways, limiting the effectiveness of conventional statistical models such as logistic regression, which typically assume independence among predictors and linear relationships \cite{4}. While such models are valued for their interpretability, they often fall short in handling high-dimensional, correlated, or imbalanced data. The issue of class imbalance is particularly problematic in stroke datasets, where the number of stroke-positive cases is significantly lower than non-stroke instances. This imbalance can skew model training, resulting in poor sensitivity toward the minority class, which in this context is the clinically critical outcome \cite{5}. Additionally, real-world healthcare data is often plagued by inconsistencies such as missing values, measurement errors, and institutional variability in data collection practices, all of which undermine the robustness and generalizability of predictive models. Compounding these issues is the limited interpretability of many complex models, which, although potentially accurate, fail to gain clinical traction due to their opaque decision-making processes and lack of transparency \cite{6}. An equally important but frequently overlooked issue is feature selection; including irrelevant or noisy variables can lead to overfitting and reduced performance in practical deployment. To address these challenges, there is a pressing need for predictive frameworks that are not only methodologically sound but also clinically aligned—models that incorporate thorough data preprocessing, manage class imbalance effectively, and employ structured, justifiable feature selection strategies. Such frameworks must prioritize interpretability and reliability to ensure integration into diverse clinical environments and support meaningful, preventive healthcare interventions.

Over the past five years, a growing number of studies have explored data-driven methods for predicting stroke risk, applying a variety of machine learning models including decision trees, support vector machines, and neural networks. While these approaches have shown moderate success, many are constrained by the use of a single classifier, limiting the ability to benchmark performance across diverse modeling strategies \cite{7}. This narrow scope hampers the understanding of which algorithms are most effective under different clinical data conditions, particularly in the presence of noisy, imbalanced, or high-dimensional datasets. Ensemble learning techniques—such as Random Forest, Gradient Boosting, and Stacking—offer a promising alternative by combining the predictive strengths of multiple models, thereby improving generalizability and reducing overfitting. However, their adoption in stroke prediction research remains limited, with few studies rigorously evaluating their performance in comparison to individual classifiers \cite{8}. Another critical yet often overlooked aspect of existing research is feature selection. While automated techniques such as LASSO regression or tree-based importance ranking have been used in some cases, many studies either omit this step or rely on manually selected variables without sufficient justification. This lack of methodological transparency not only weakens reproducibility but also diminishes the interpretability and clinical relevance of the models. Furthermore, the well-documented problem of class imbalance in stroke datasets is frequently addressed inadequately. Techniques like SMOTE (Synthetic Minority Oversampling Technique), which have been shown to be effective in related fields such as cardiovascular disease and diabetes, are inconsistently applied in stroke modeling \cite{9}. Evaluation metrics also vary widely, with many studies reporting only overall accuracy—a measure that can be misleading in imbalanced scenarios. Essential metrics such as F1-score, area under the ROC curve (AUC), precision-recall curves, and Matthews Correlation Coefficient (MCC) are often omitted, obscuring a complete understanding of model performance \cite{10}. Collectively, these limitations highlight the need for more rigorous, transparent, and methodologically robust frameworks that integrate ensemble learning, validated feature selection, and comprehensive performance evaluation for clinically actionable stroke prediction.

This study proposes a structured and clinically aligned framework for early stroke risk prediction, addressing the critical gaps in interpretability, model robustness, and methodological rigor identified in previous research. The framework integrates comprehensive data preprocessing, advanced feature selection, and ensemble-based classification strategies, with the goal of developing a system that is both accurate and applicable in real-world clinical settings. Key Contributions of this study include,

\begin{itemize}
    \item A complete data preprocessing pipeline that addresses missing values, outliers, and class imbalance using SMOTE, ensuring data quality and balance.
    \item Integration of advanced feature selection methods (correlation filtering, tree-based ranking) to enhance model interpretability and reduce dimensionality.
    \item Implementation and comparison of multiple baseline classifiers and ensemble models to identify optimal predictive structures using hyperparameter tuning with GridSearchCV for three feature sets.
    \item Proposal of an ensemble model that outperforms traditional models in terms of predictive performance and robustness.
\end{itemize}

By combining methodological precision with clinical relevance, this study contributes a reproducible and interpretable predictive framework tailored for stroke risk assessment. The findings aim to support clinicians in identifying high-risk individuals earlier, facilitating timely intervention and improving long-term patient outcomes.

\section{Methodology}
\begin{figure}[h]
    \centering
    \includegraphics[width=\linewidth, height=0.75\textheight, keepaspectratio]{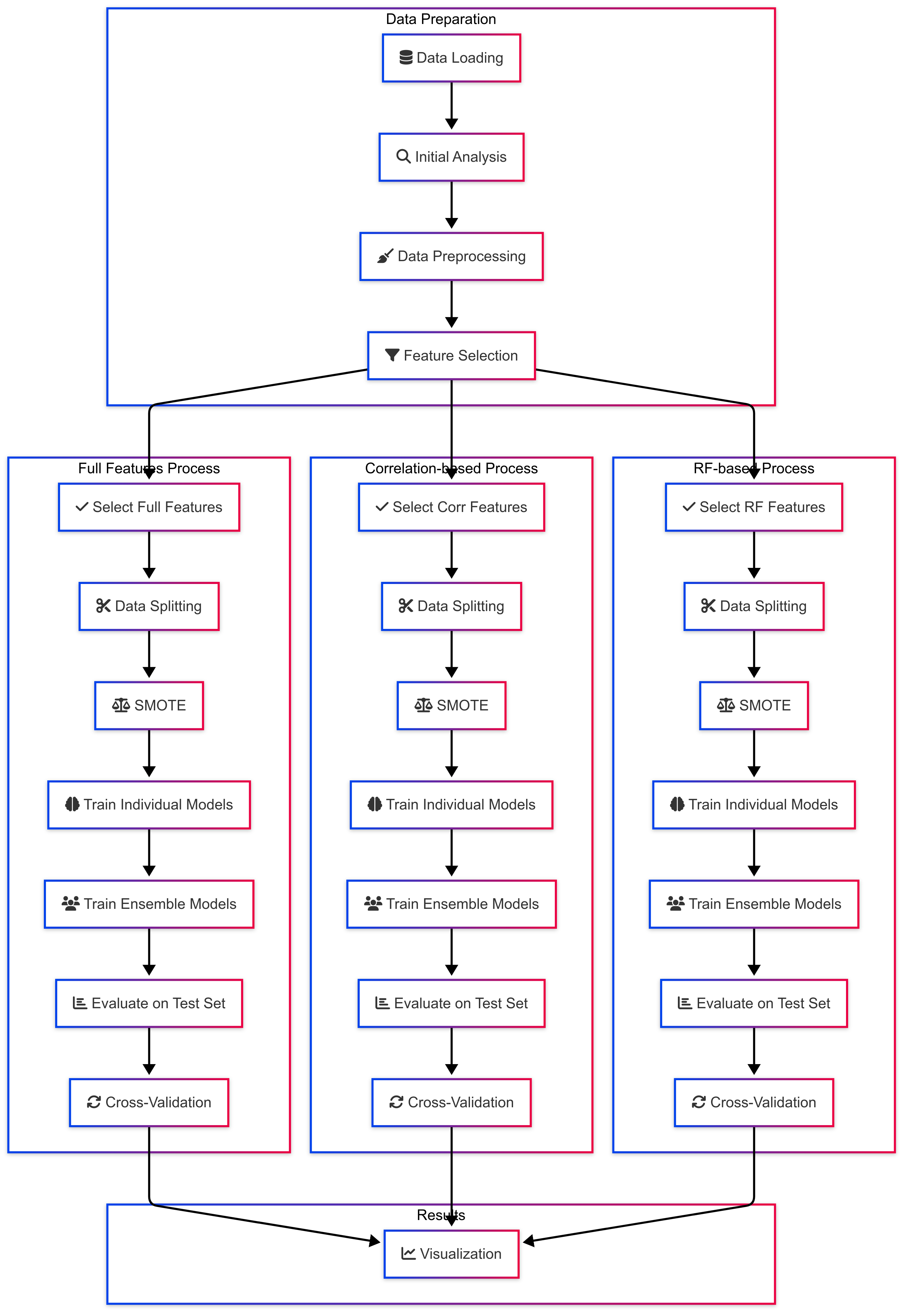} % Adjust the factor as needed
    \caption{A visual overview of the key steps involved in this study}
    \label{fig:visual-overview-in-this-study}
\end{figure}

This study proposes a methodological framework for predicting the occurrence of a brain stroke employing machine learning algorithms. The approach involves comprehensive preprocessing of the data, followed by feature engineering and comparison of  several classification models to determine the best predictive strategies. Therefore, this study has helped develop a model that can be helpful to the stroke care industry to provide a better system for predicting strokes in the community and potentially decrease its related morbidity and mortality. The global process is summarized in Figure~\ref{fig:visual-overview-in-this-study}, showing the process used in this study from data preparation to model evaluation.

\subsection{Dataset Acquisition}
This study was performed using a complete brain stroke dataset consisting of 4,981 patient records with 11 features, including the demographical information, medical history, and lifestyle factors \cite{11}. The dataset contains records of several potential risk factors for stroke and has the target variable \texttt{stroke}, which is encoded as binary (1 if stroke occurred, 0 if not). This database contains categorical variables (\texttt{gender}, \texttt{ever\_married}, \texttt{work\_type}, \texttt{Residence\_type}, \texttt{smoking\_status}), numerical variables (\texttt{age}, \texttt{avg\_glucose\_level}, \texttt{bmi}), and the given binary indicators for \texttt{hypertension} and \texttt{heart\_disease}. 
An initial quality evaluation verified that the dataset is intact with no absent values, eliminating the need for imputation or data augmentation methods. This bolsters the dependability of the following analyses. The overview of the dataset configuration is displayed in Table~\ref{tab:dataset-overview}..

\begin{table*}[htbp]
\centering
\caption{Dataset Overview}
\vspace{0.5em}
\label{tab:dataset-overview}
\begin{tabular}{|c|l|l|l|}
\hline
\textbf{No.} & \textbf{Feature} & \textbf{Data Type} & \textbf{Description} \\
\hline
1  & \texttt{gender}               & Categorical & Biological sex of the patient \\
2  & \texttt{age}                  & Numerical   & Age in years \\
3  & \texttt{hypertension}         & Binary      & 1 if patient has hypertension \\
4  & \texttt{heart\_disease}       & Binary      & 1 if patient has heart disease \\
5  & \texttt{ever\_married}        & Categorical & Marital status \\
6  & \texttt{work\_type}           & Categorical & Type of employment \\
7  & \texttt{Residence\_type}      & Categorical & Urban or rural residence \\
8  & \texttt{avg\_glucose\_level}  & Numerical   & Average glucose level \\
9  & \texttt{bmi}                  & Numerical   & Body Mass Index \\
10 & \texttt{smoking\_status}      & Categorical & Smoking behavior \\
11 & \texttt{stroke}               & Binary      & Stroke occurrence (target variable) \\
\hline
\end{tabular}
\end{table*}

\subsection{Exploratory Data Analysis (EDA)}

\subsubsection{Class Distribution Analysis}
During exploratory analysis, one of the basic problems identified was the high class imbalance in the target. This imbalance can be seen in Figure 2, which shows that stroke cases made a small minority of our total number of cases and only accounted for approximately 5\% of our overall dataset.

\begin{figure}[h]
\centering
\begin{tikzpicture}
\begin{axis}[
    ylabel=Count,
    xlabel=Stroke Status,
    enlargelimits=0.1,
    legend style={at={(0.95,0.95)}, anchor=north east},
    ybar,
    bar width=35pt,
    xtick={1,2},
    xticklabels={0, 1},
    ymin=0,
    enlarge x limits=1,
    axis on top=false,
    nodes near coords,
    nodes near coords align={vertical},
    point meta=explicit symbolic,
]
\addplot[fill=blue] coordinates {(1,4733) [95.02\%]};
\addplot[fill=red] coordinates {(2,248) [4.98\%]};
\legend{No Stroke, Stroke}
\end{axis}
\end{tikzpicture}
\caption{Class Distribution of Stroke Status}
\label{fig:stroke_distribution}
\end{figure}
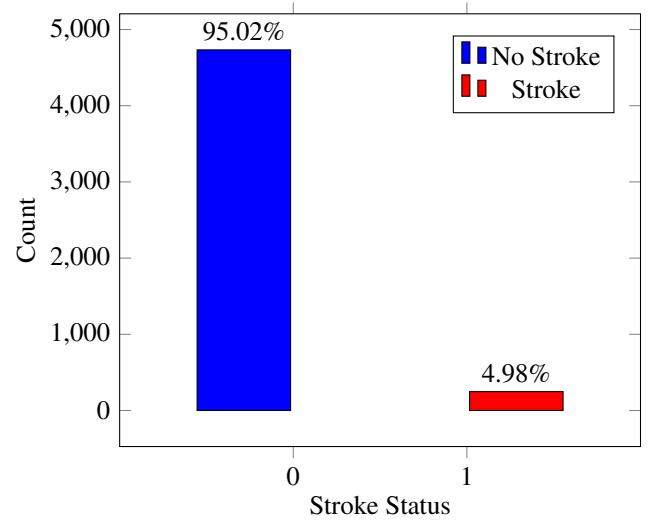

We can represent this imbalance mathematically in the class proportion: 
\( P(y = k) = \frac{n_k}{N} \), where \( n_k \) is the number of instances that belong to class \( k \), and \( N \) is the total number of instances. 
For stroke cases, \( P(y = 1) = 0.05 \), while for non-stroke cases, \( P(y = 0) = 0.95 \).

\subsubsection{Numerical Feature Analysis}

We carried out a distribution analysis of each numerical feature (\textit{age}, \textit{avg\_glucose\_level}, and \textit{bmi}) through descriptive statistics and visualization. Central tendency measures, dispersion, and shape parameters (skewness and kurtosis) were calculated to characterize the distributions.

Skewness, a measure of the distribution asymmetry, was calculated using:

\begin{equation}
\text{Skewness} = \frac{1}{n} \sum_{i=1}^{n} \left( \frac{x_i - \bar{x}}{s} \right)^3
\end{equation}

Kurtosis, which captures the "tailedness" of a distribution, was computed as:

\begin{equation}
\text{Kurtosis} = \frac{1}{n} \sum_{i=1}^{n} \left( \frac{x_i - \bar{x}}{s} \right)^4 - 3
\end{equation}

where \( x_i \) represents individual data points, \( \bar{x} \) is the sample mean, \( s \) is the standard deviation, and \( n \) is the total number of data points.

\begin{figure}[htbp]
    \centering
    \includegraphics[width=\linewidth]{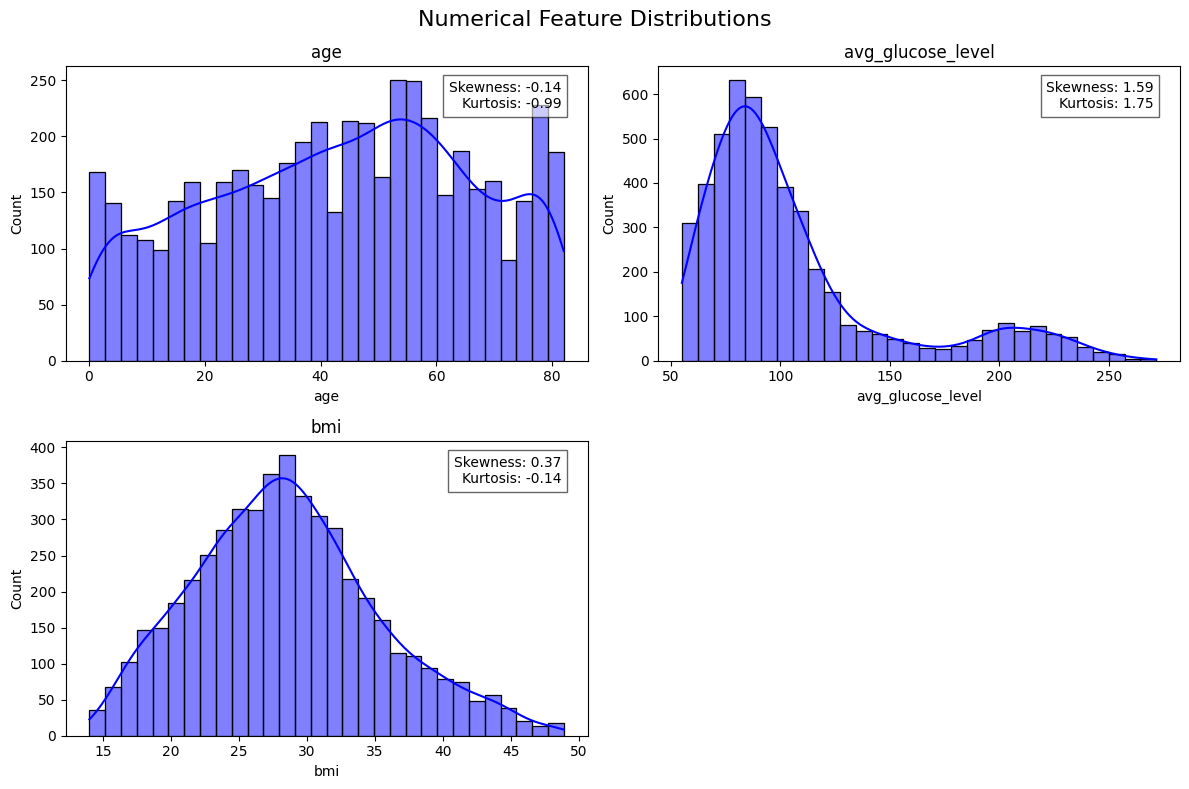} % replace with your file name
    \caption{Numerical Feature Distributions}
    \label{fig:stroke-overview}
\end{figure}

The analysis revealed distinct distribution patterns for each numerical feature.The age distribution followed a relatively normal pattern with minimal skewness ($\approx -0.14$) and moderate kurtosis ($\approx -0.99$), indicating a slightly platykurtic (flatter than normal) shape. The age range was broad, with peaks observed in middle-aged and elderly populations, aligning with the epidemiological profile that stroke risk increases with age.The average glucose level exhibited a pronounced right-skewed distribution (skewness $\approx 1.59$) with high kurtosis ($\approx 1.75$), signifying numerous outliers in the upper tail. This pattern suggests that while most patients had glucose levels within normal ranges, a significant subset presented with elevated levels, which may contribute to increased stroke risk.
BMI followed a moderately right-skewed distribution (skewness $\approx 0.37$), with minor positive kurtosis ($\approx -0.14$), and a broad range of values with a majority in the overweight and obese ranges. This breakdown correlates with the established clinical knowledge that higher BMI is a risk factor for cardiovascular events such as stroke. Visualizing these distributions with histograms, complemented by kernel density estimates, revealed important insights about the data and its potential preprocessing needs, especially related to outliers.

\subsubsection{Categorical Feature Analysis}

The composition of categorical variables in the sample population and the relationship between the categorical variables and stroke were evaluated visually via frequency distribution \cite{12}. For each categorical feature, a count plot was created, and proportions were calculated to identify dominant categories and detect any imbalance within the respective feature.

The distribution ratio for every category was determined as:

\begin{equation}
P(X = x_j) = \frac{\text{count}(X = x_j)}{N}
\end{equation}

where \( X \) represents the categorical variable, and \( x_j \) is a specific category within that variable. The numerator, \( \text{count}(X = x_j) \), refers to the number of instances in the dataset where the categorical variable \( X \) takes the value \( x_j \). The denominator, \( N \), represents the total number of instances in the dataset. These proportions help evaluate how the dataset is distributed across different categories in a feature. Based on this information, data scientists can identify potential imbalances and biases in the dataset that may affect downstream analysis or model performance.

\begin{figure*}[htbp]
    \centering
    \includegraphics[width=\linewidth]{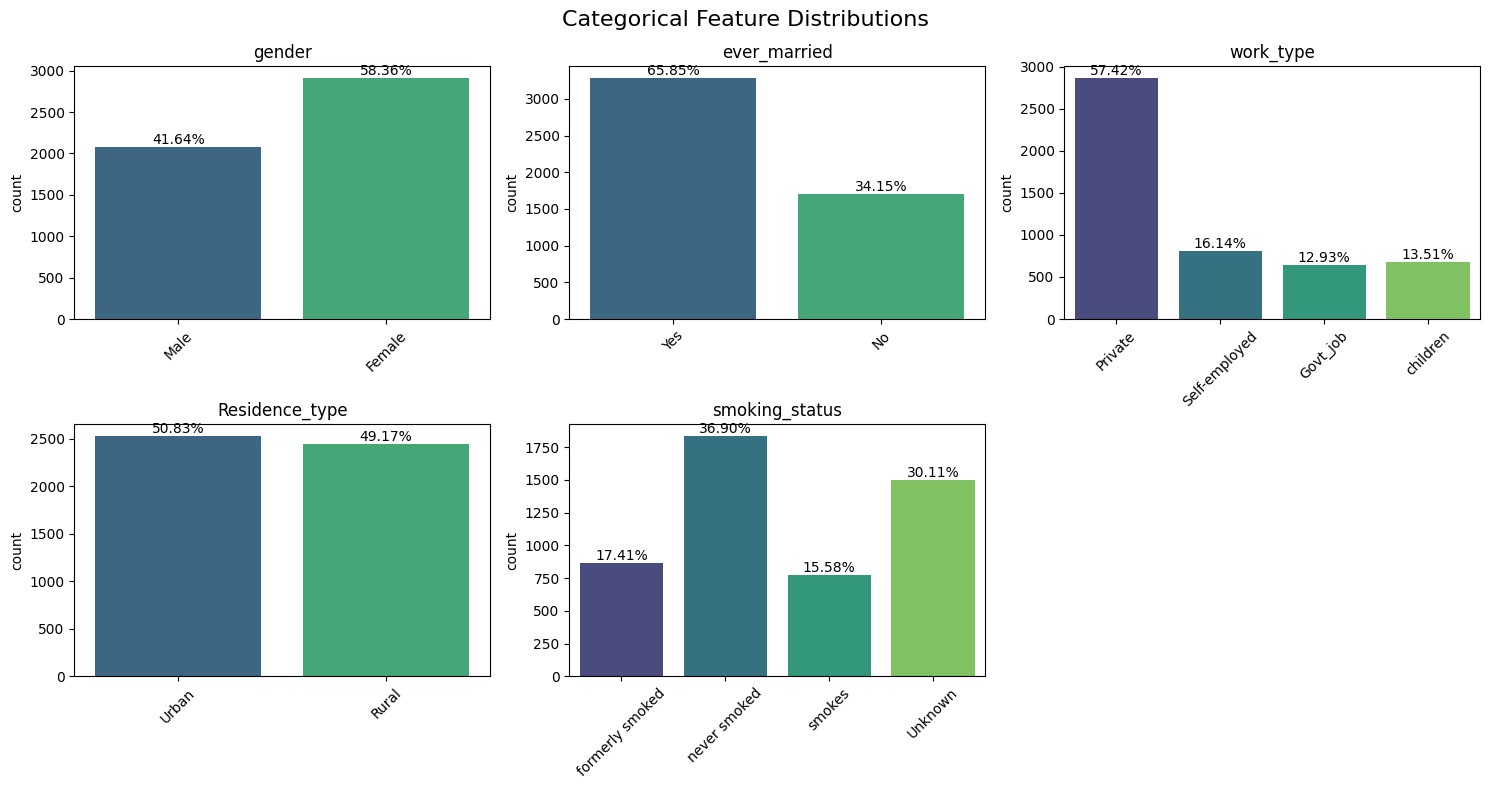}
    \caption{Categorical Feature Distributions}
    \label{fig:stroke-overview}
\end{figure*}

The analysis of categorical features revealed several important patterns. The gender distribution showed that females comprised approximately 58.36\% of the dataset, while males made up 41.64\%, indicating a slight imbalance in gender representation. Regarding marital status, the majority of individuals were married (65.85\%), whereas 34.15\% were unmarried, which aligns with typical age distributions, as older populations are more likely to be married.

The work type distribution indicated that \textit{Private} employment was the most common category (57.42\%), followed by \textit{Self-employed} (16.14\%), \textit{Government job} (12.93\%), and \textit{Children} (13.51\%). This distribution reflects common employment patterns among adults. The residence type feature was fairly balanced, with Urban residents comprising 50.83\% and Rural residents 49.17\%, ensuring a well-represented sample for residential comparisons.

Smoking status revealed that \textit{Never smoked} was the most common category (36.90\%), followed by \textit{Unknown} (30.11\%), \textit{Formerly smoked} (17.41\%), and \textit{Smokes} (15.58\%). The substantial proportion of \textit{Unknown} responses may present a limitation, as smoking is a known risk factor for stroke.

These categorical distributions provide valuable insights into the demographic and lifestyle characteristics of the patient population, while also highlighting areas where feature engineering or stratified analysis might be beneficial.

\subsection{Data Preprocessing Pipeline}

\subsubsection{Categorical Feature Encoding}
Categorical variables were transformed into numerical representations to make them compatible with machine learning algorithms \cite{13}. We employed \texttt{LabelEncoder} from the \texttt{scikit-learn} library, which assigns an integer value to each unique category.

For a categorical variable \( X \) with \( k \) unique categories, the encoding function \( f \) maps each category to an integer as follows:

\begin{equation}
f: \{ c_1, c_2, \ldots, c_k \} \rightarrow \{ 0, 1, \ldots, k-1 \}
\end{equation}

The encoding mappings were carefully preserved to maintain interpretability, and the mappings are presented in Table~\ref{tab:label_encoding}.

\begin{table}[h]
  \caption{Label encoding scheme for categorical features}
  \vspace{0.5em}
  \centering
  \begin{tabular}{|l|l|c|}
    \hline
    \textbf{Feature} & \textbf{Category}     & \textbf{Encoded Value} \\ \hline
    \multirow{2}{*}{gender}          & Female           & 0 \\ \cline{2-3}
                                     & Male             & 1 \\ \hline
    \multirow{2}{*}{ever\_married}   & No               & 0 \\ \cline{2-3}
                                     & Yes              & 1 \\ \hline
    \multirow{4}{*}{work\_type}      & Govt\_job        & 0 \\ \cline{2-3}
                                     & Private          & 1 \\ \cline{2-3}
                                     & Self-employed    & 2 \\ \cline{2-3}
                                     & children         & 3 \\ \hline
    \multirow{2}{*}{Residence\_type} & Rural            & 0 \\ \cline{2-3}
                                     & Urban            & 1 \\ \hline
    \multirow{4}{*}{smoking\_status} & Unknown          & 0 \\ \cline{2-3}
                                     & formerly smoked  & 1 \\ \cline{2-3}
                                     & never smoked     & 2 \\ \cline{2-3}
                                     & smokes           & 3 \\ \hline
  \end{tabular}
  \label{tab:label_encoding}
\end{table}

This encoding approach maintains the ordinal relationship between categories where appropriate — for example, in \textit{smoking\_status}, where \textit{never smoked}, \textit{formerly smoked}, and \textit{smokes} represent increasing levels of exposure. For truly nominal variables without inherent ordering, such as \textit{work\_type}, the assigned numerical values are arbitrary but consistent throughout the analysis.While one-hot encoding is often preferred for nominal categorical variables to avoid imposing artificial ordering, our preliminary experiments showed that \texttt{LabelEncoder} provided comparable performance while resulting in more computationally efficient models due to the reduced dimensionality. 
This trade-off was considered acceptable given the relatively small number of categories in each feature.

\subsubsection{Outlier Detection and Removal}

Outliers can significantly impact model performance, particularly for algorithms sensitive to extreme values. We implemented the Interquartile Range (IQR) method to identify and remove outliers from numerical features \cite{14}. This robust statistical approach defines boundaries based on quartile distribution, which is less sensitive to extreme values compared to methods based on standard deviation.

For each numerical feature, we calculated:
\begin{align*}
Q1 &= \text{25th percentile} \\
Q3 &= \text{75th percentile} \\
\end{align*}
\begin{equation}
\text{IQR} = Q3 - Q1
\end{equation}

Lower and upper boundaries were established using:
\begin{equation}
\text{Lower Bound} = Q1 - 1.5 \times \text{IQR}
\end{equation}
\begin{equation}
\text{Upper Bound} = Q3 + 1.5 \times \text{IQR}
\end{equation}

Any data points falling outside these boundaries were considered outliers and removed from the dataset. The detailed results of outlier detection for each numerical feature are shown in Table~\ref{tab:outlier_detection}.

\begin{table*}[h]
\caption{Outlier Detection Summary Using IQR Method.}
\vspace{0.5em}
\centering
\begin{tabular}{|l|c|c|c|c|c|}
\hline
\textbf{Feature} & \textbf{Q1} & \textbf{Q3} & \textbf{Lower Bound} & \textbf{Upper Bound} & \textbf{Number of Outliers} \\
\hline
age & 25.00 & 61.00 & -29.00 & 115.00 & 0 \\
\hline
avg\_glucose\_level & 77.23 & 113.86 & 22.29 & 168.81 & 602 \\
\hline
bmi & 23.20 & 32.00 & 10.00 & 45.20 & 42 \\
\hline
\end{tabular}
\label{tab:outlier_detection}
\end{table*}

This analysis revealed interesting patterns. No outliers were detected in \textit{age}, suggesting that all age values fell within the expected range for a population-based stroke study. A substantial number of outliers (602, approximately 12\% of the dataset) were identified in the \textit{avg\_glucose\_level}, primarily in the upper range. These likely represent patients with severe hyperglycemia or uncontrolled diabetes — conditions known to increase stroke risk. A smaller number of outliers (42, approximately 0.8\% of the dataset) were found in \textit{bmi}, representing individuals with extreme underweight or obesity.
After outlier removal, the dataset was reduced from 4,981 to 4,337 records, representing a 13\% reduction in dataset size. While this reduction is substantial, it results in a more homogeneous and statistically reliable dataset for model development.

\begin{figure}[h]
    \centering
    \includegraphics[width=\linewidth]{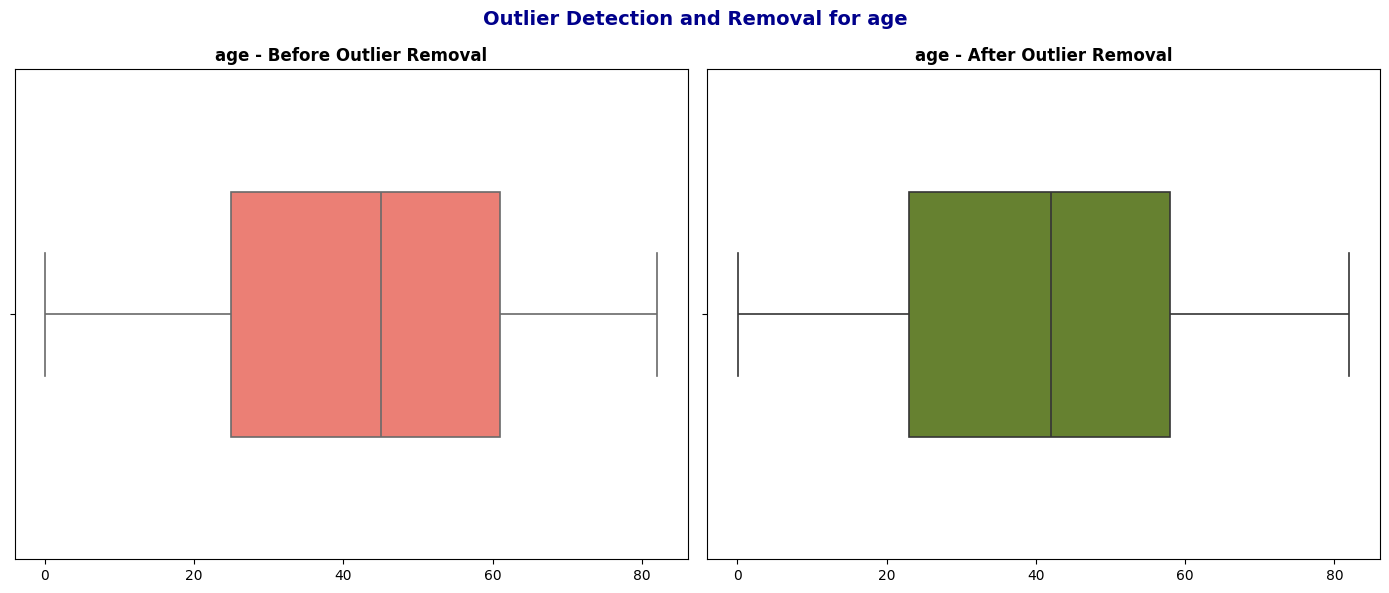} \\
    \includegraphics[width=\linewidth]{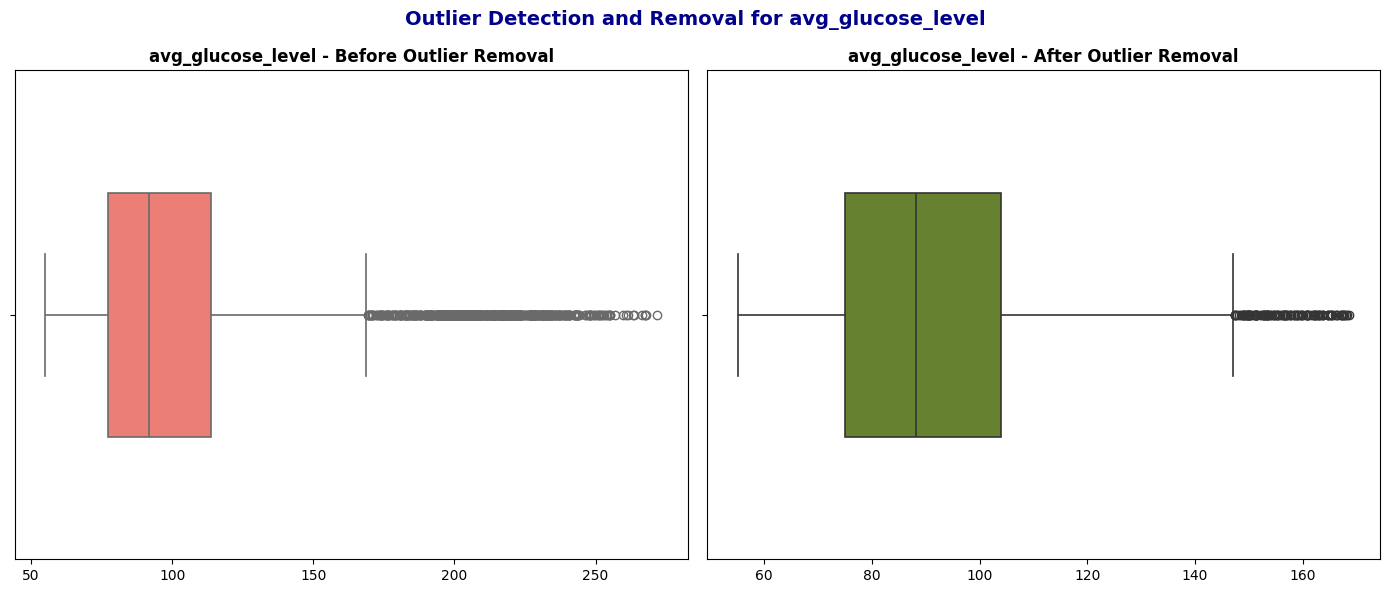} \\
    \includegraphics[width=\linewidth]{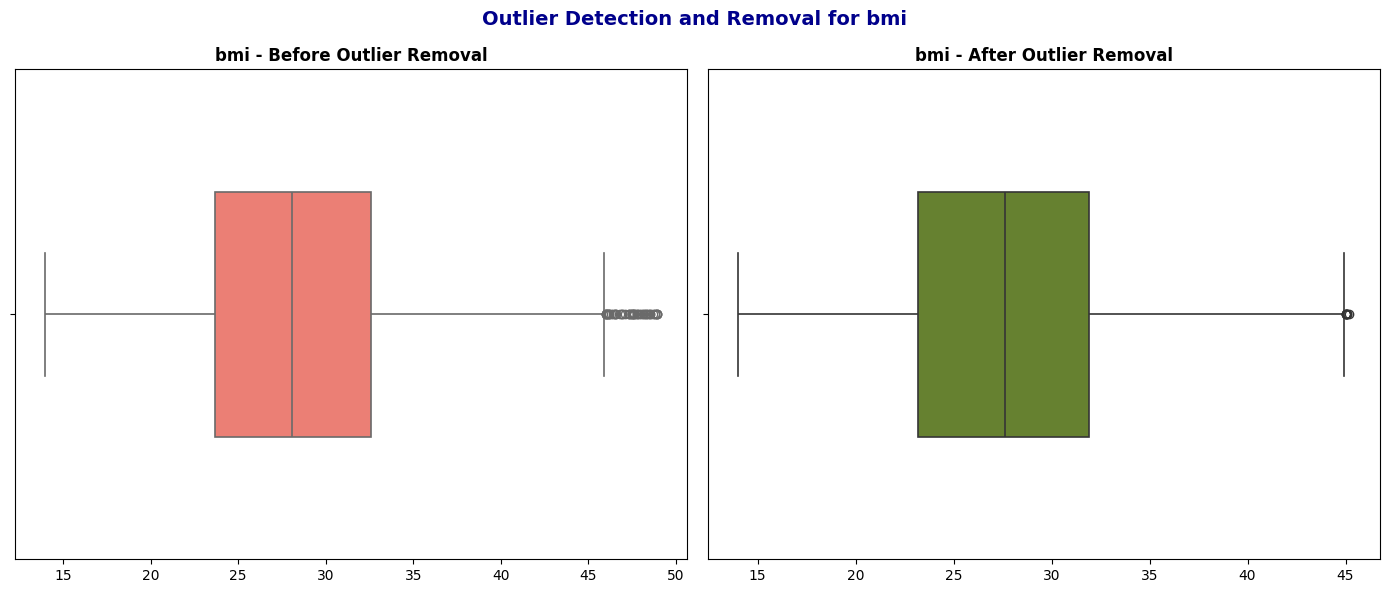}
    \caption{Box Plots Before and After Outlier Removal}
    \label{fig:boxplots}
\end{figure}

The box plots in Figure ~\ref{fig:boxplots} visually demonstrate the impact of outlier removal on data distribution. The most dramatic change is observed in \textit{avg\_glucose\_level}, where the removal of extreme values resulted in a more compact distribution with a significantly reduced upper range. This preprocessing step enhances the dataset's suitability for modeling by reducing the influence of extreme values that could potentially bias learning algorithms toward rare, extreme cases rather than capturing general patterns of stroke risk.

\subsection{Feature Selection method}

Feature selection is crucial for developing interpretable and efficient machine learning models. We implemented two complementary approaches to identify the most predictive variables for stroke prediction, capturing both linear and non-linear relationships with the target variable.

\subsubsection{Correlation-Based Feature Selection}
Pearson correlation coefficients were calculated between each feature and the target variable (\textit{stroke}). For continuous variables, the Pearson correlation $\rho$ coefficient was calculated as:

\begin{equation}
\rho_{X,Y} = \frac{\text{cov}(X,Y)}{\sigma_X \sigma_Y} = \frac{\mathbb{E}[(X - \mu_X)(Y - \mu_Y)]}{\sigma_X \sigma_Y}
\end{equation}

Where $\text{cov}(X,Y)$ is the covariance between variables $X$ and $Y$, $\sigma_X$ and $\sigma_Y$ are their standard deviations, $\mu_X$ and $\mu_Y$ are their means, and $\mathbb{E}$ is the expectation operator \cite{15}. For categorical variables, the point-biserial correlation was calculated as a special case of the Pearson correlation when one variable is dichotomous.

\begin{figure}[h]
\centering
 \includegraphics[width=\linewidth]{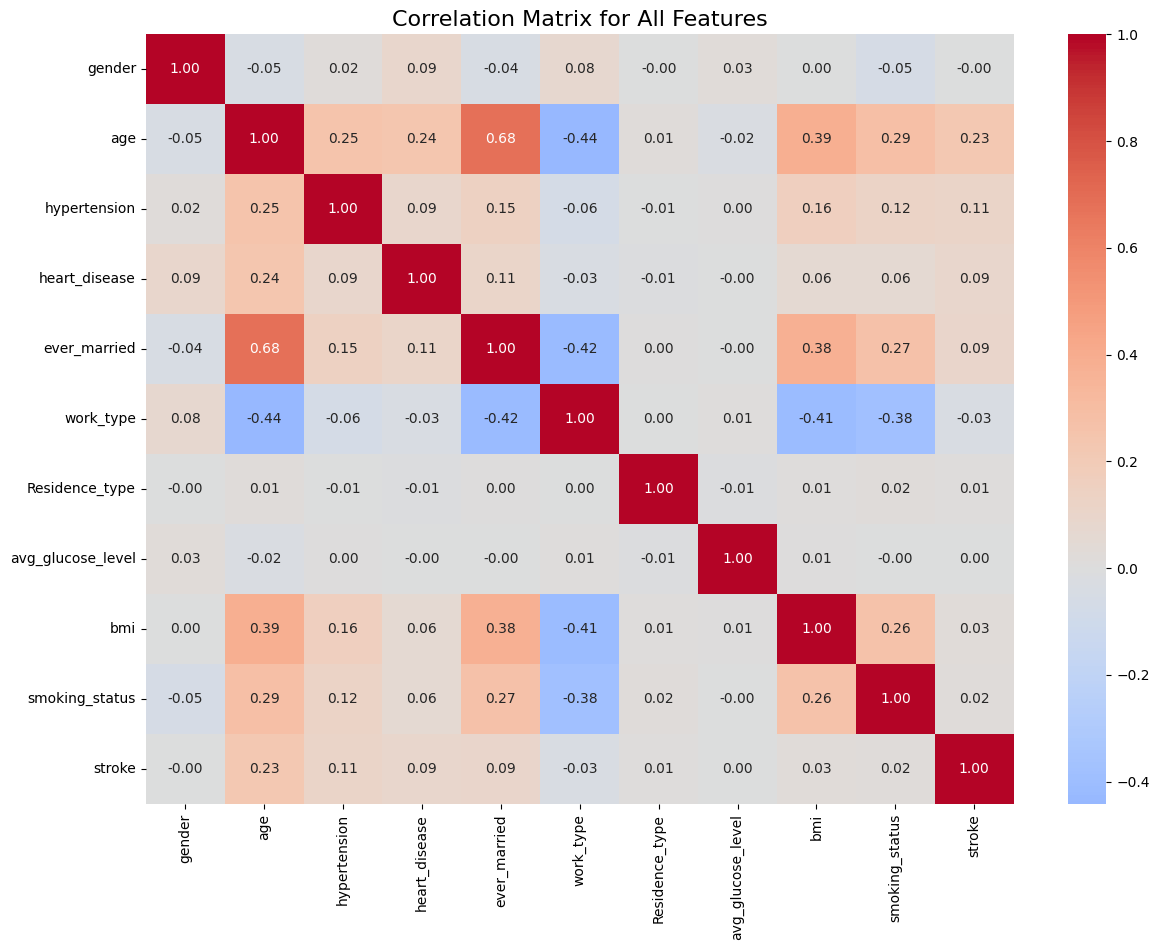} % 
\caption{Correlation Matrix Heatmap}
\label{fig:correlation_heatmap}
\end{figure}

Figure~\ref{fig:correlation_heatmap} presents the correlation matrix heatmap, visualizing relationships between all features in the dataset and revealing varying strengths of correlation with stroke occurrence. \textit{Age} exhibited the strongest correlation (0.23), reinforcing its critical role in stroke risk. Moderate correlations were observed for \textit{hypertension} (0.11), \textit{ever\_married} (0.09), and \textit{heart\_disease} (0.09), which are well-established stroke risk factors. Weaker correlations were found for \textit{work\_type} (-0.03), \textit{BMI} (0.03), and \textit{smoking\_status} (0.02), while very weak correlations were noted for \textit{residence\_type} (0.01), \textit{gender} (-0.00), and \textit{average\_glucose\_level} (0.03). Based on a selection threshold of 0.02, the significant predictors included \textit{age}, \textit{hypertension}, \textit{ever\_married}, \textit{heart\_disease}, \textit{BMI}, and \textit{smoking\_status}.

The correlation matrix also revealed notable inter-feature relationships that provide additional insights into stroke risk factors. \textit{Age} and \textit{ever\_married} showed a strong positive correlation (0.68), indicating that older individuals were more likely to be married. \textit{Hypertension} and \textit{heart\_disease} had a weaker correlation (0.09), suggesting a less direct relationship than initially expected. Additionally, \textit{age} and \textit{hypertension} demonstrated a positive correlation (0.25), aligning with clinical observations that hypertension prevalence increases with age.

\subsubsection{Random Forest-Based Feature Selection}
Correlation analysis primarily captures linear relationships between variables. To identify potentially non-linear relationships and interaction effects, we employed a Random Forest classifier to determine feature importance based on the Gini impurity reduction \cite{16}. For a feature $X_j$, its importance $I(X_j)$ was calculated as:

\begin{equation}
I(X_j) = \sum_{t \in T} p(t) \cdot i(t, X_j)
\end{equation}

Where $T$ is the set of all trees in the forest, $p(t)$ is the proportion of samples reaching node $t$, and $i(t, X_j)$ is the decrease in impurity at node $t$ due to feature $X_j$. The Gini impurity at a node $t$ is defined as:

\begin{equation}
G(t) = \sum_{k} p_{t,k}(1 - p_{t,k}) = 1 - \sum_{k} p_{t,k}^2
\end{equation}
Where $p_{t,k}$ is the proportion of samples at node $t$ that belong to class $k$.

\begin{figure}[h]
\centering
 \includegraphics[width=\linewidth]{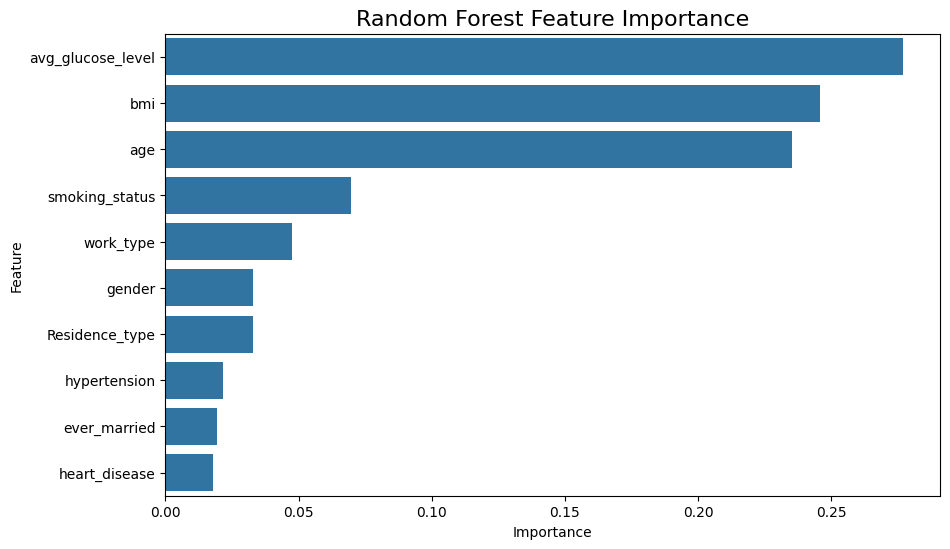} % 
\caption{Random Forest Feature Importance}
\label{fig:rf_feature_importance}
\end{figure}

Figure~\ref{fig:rf_feature_importance} presents the Random Forest feature importance ranking, offering a complementary perspective compared to the correlation analysis. This method identified \textit{average glucose level} ($\approx 0.27$), \textit{BMI} ($\approx 0.24$), and \textit{age} ($\approx 0.23$) as the most critical predictors, highlighting their strong contribution to stroke prediction. Moderate importance was attributed to \textit{smoking status} ($\approx 0.08$), \textit{work type} ($\approx 0.07$), \textit{gender} ($\approx 0.05$), and \textit{residence type} ($\approx 0.04$), indicating these factors still played a meaningful role in model decisions. Lower importance was observed for \textit{hypertension} ($\approx 0.015$), \textit{heart disease} ($\approx 0.013$), and \textit{ever married} ($\approx 0.014$), suggesting a relatively smaller influence in the model's decision-making process.

Using a feature importance threshold of $0.025$, the key predictors identified were \textit{average glucose level}, \textit{BMI}, \textit{age}, \textit{smoking status}, \textit{work type}, \textit{gender}, and \textit{residence type}. This non-linear feature importance ranking provided deeper insights into stroke risk factors, capturing complex interactions that correlation analysis alone might overlook.

\subsubsection{Analysis of Selected Features}
The two feature selection approaches provided complementary perspectives on feature relevance, as illustrated in Table~\ref{tab:feature_comparison}. This two feature selection approaches provided complementary perspectives on feature relevance, as illustrated in Table~\ref{tab:feature_comparison}. This comparison revealed interesting patterns. Age, BMI, work type, and smoking status were identified as important by both methods, suggesting their robust predictive power across different analytical techniques.

\begin{table}[t]
  \caption{Comparison of Selected Features by Different Methods}
  \vspace{0.5em}
  \label{tab:feature_comparison}
  \footnotesize
  \centering
  \begin{adjustbox}{max width=\columnwidth}
    \begin{tabular}{|l|l|}
      \hline
      \textbf{Correlation Selected Features} & \textbf{Random Forest Selected Features} \\ \hline
      age            & avg\_glucose\_level \\ \hline
      hypertension   & bmi                 \\ \hline
      ever\_married  & age                 \\ \hline
      heart\_disease & smoking\_status     \\ \hline
      work\_type     & work\_type          \\ \hline
      bmi            & gender              \\ \hline
      smoking\_status& residence\_type     \\ \hline
    \end{tabular}
  \end{adjustbox}
\end{table}

Hypertension, ever\_married, and heart\_disease were highlighted only by the correlation-based method, indicating moderate linear relationships with stroke. These features may have direct associations with stroke occurrence but might not contribute significantly in a non-linear model like Random Forest.

On the other hand, average glucose level, gender, and residence type were emphasized primarily by the Random Forest approach, suggesting non-linear relationships or interaction effects with stroke occurrence. The discrepancy in the importance of \textit{avg\_glucose\_level} is particularly notable. Despite its linear correlation with stroke (0.03), which technically meets the correlation-based threshold (0.02), it was excluded from the correlation-selected features in Table~\ref{tab:feature_comparison}. However, it emerged as the most important feature in the Random Forest analysis (importance $\approx$ 0.27). This suggests that \textit{avg\_glucose\_level} may have complex, non-linear relationships with stroke occurrence that are not captured by simple correlation measures. For example, it might interact with other features like age or BMI to influence stroke risk, making it a crucial variable in predictive modeling.

\subsection{Dataset Preparation for Modeling}
To assess the impact of feature selection, we created three feature sets: \textit{Full Features} (all 10 predictors), \textit{Correlation-based Features} (7 features selected via correlation thresholding), and \textit{Random Forest-based Features} (7 features chosen based on tree-based importance). Each set was split into training (80\%) and testing (20\%) using stratified sampling to preserve stroke prevalence ($\approx$5\%). The train-test split was performed using \texttt{train\_test\_split(X, y, test\_size=0.2, random\_state=42, stratify=y)}, ensuring balanced class distribution for unbiased model evaluation, which is demonstrated in Table~\ref{tab:dataset-dimensions}.

\begin{table}[t]
  \caption{Dataset Dimensions After Train–Test Split}
  \vspace{0.5em}
  \label{tab:dataset-dimensions}
  \footnotesize
  \centering
  \begin{adjustbox}{max width=\columnwidth}
    \begin{tabular}{|l|c|c|}
      \hline
      \textbf{Feature Set} & \textbf{Training Set} & \textbf{Testing Set} \\ \hline
      Full Features               & (3469, 10) & (868, 10) \\ \hline
      Correlation‑based Features  & (3469, 7)  & (868, 7)  \\ \hline
      Random Forest‑based Features& (3469, 7)  & (868, 7)  \\ \hline
    \end{tabular}
  \end{adjustbox}
\end{table}

\subsubsection{Standardization }
Numerical features were standardized using \texttt{StandardScaler} to normalize scale differences \cite{17}, ensuring uniform feature distribution and preventing dominance by variables with larger ranges, such as \textit{avg\_glucose\_level} (ranging from approximately 55 to 271). Standardization was performed using:

\begin{equation}
X_{\text{scaled}} = \frac{X - \mu}{\sigma}
\end{equation}

where $\mu$ and $\sigma$ represent the mean and standard deviation of each feature from the training set. Rescaling ensures models can efficiently converge to the optimal solution space, which is particularly beneficial for distance-based models such as Support Vector Machines (SVMs) and k-Nearest Neighbors (KNN), as well as for gradient-based optimizers used in neural networks and some ensemble methods.

However, standardization has minimal impact on tree-based models such as Random Forest and Gradient Boosting, which are invariant to monotonic transformations. Normalizing the dataset ensures that each variable contributes equally to distance calculations, facilitating faster training and improved performance across a broad range of machine learning models.

\subsection{Addressing Class Imbalance}

\subsubsection{SMOTE Implementation and Theoretical Foundation}

The dataset exhibited a highly imbalanced class distribution, with only 5\% of instances labeled as stroke patients and 95\% as non-stroke patients. This imbalance posed a significant challenge for machine learning models, often biasing predictions toward the majority class. To address this issue, we applied SMOTE (Synthetic Minority Over-sampling Technique), which generates new synthetic instances using interpolation rather than simple replication.

SMOTE works by identifying the $k$-nearest neighbors (with $k = 5$) for each minority class instance and generating a new synthetic sample as follows:

\begin{equation}
x_{\text{new}} = x_i + \lambda (x_{\text{nn}} - x_i)
\end{equation}

where $x_i$ is a minority instance, $x_{\text{nn}}$ is one of its $k$-nearest neighbors, and $\lambda$ is a random number in the range $[0, 1]$ that determines the position of the synthetic point along the line segment between $x_i$ and $x_{\text{nn}}$ \cite{18}. This strategy improves model generalizability by generating plausible synthetic data points while minimizing overfitting. In contrast to traditional oversampling, SMOTE creates counterexamples based on diversifying the feature space, which could make the classifier more robust and improve its predictive performance.

\subsubsection{Class Distribution Transformation}
We applied SMOTE to the training data only, ensuring that the test data remained representative of the real-world class distribution. The application of SMOTE significantly altered the class distribution in the training set, as shown in Table~\ref{tab:smote-distribution}.

\begin{table}[t]
  \caption{Class Distribution Before and After SMOTE Application in Training Data}
  \vspace{0.5em}
  \label{tab:smote-distribution}
  \footnotesize             % keep it compact
  \centering
  \begin{adjustbox}{max width=\columnwidth}
    \begin{tabular}{|l|c|c|}
      \hline
      \textbf{Class} & \textbf{Before SMOTE} & \textbf{After SMOTE} \\ \hline
      No Stroke (0)  & 3,312 (95.5\%) & 3,312 (50\%) \\ \hline
      Stroke (1)     &   157 (4.5\%) & 3,312 (50\%) \\ \hline
      \textbf{Total} & 3,469 (100\%) & 6,624 (100\%) \\ \hline
    \end{tabular}
  \end{adjustbox}
\end{table}

\begin{figure}[h]
    \centering
    \includegraphics[width=\linewidth]{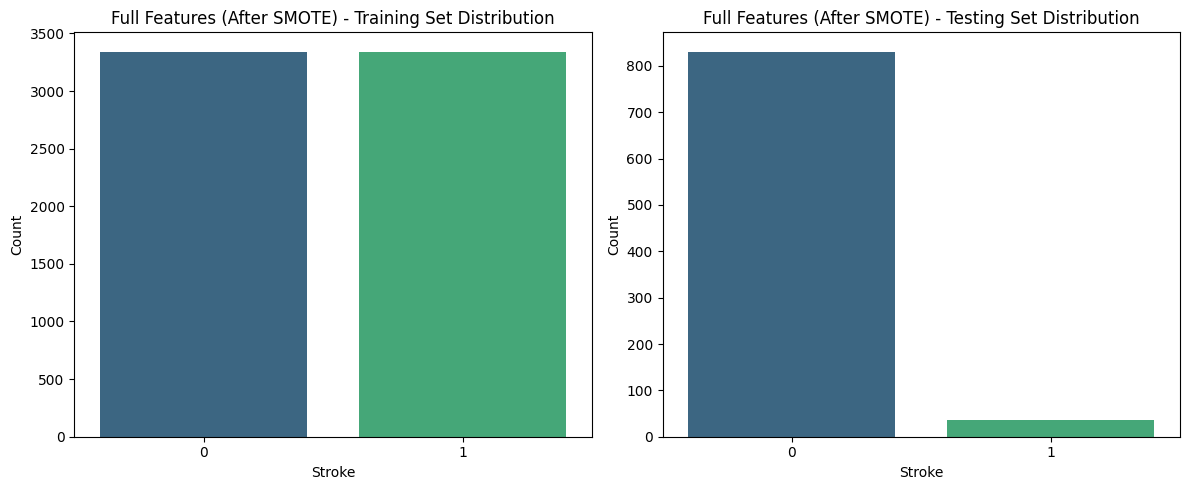} \\
    \includegraphics[width=\linewidth]{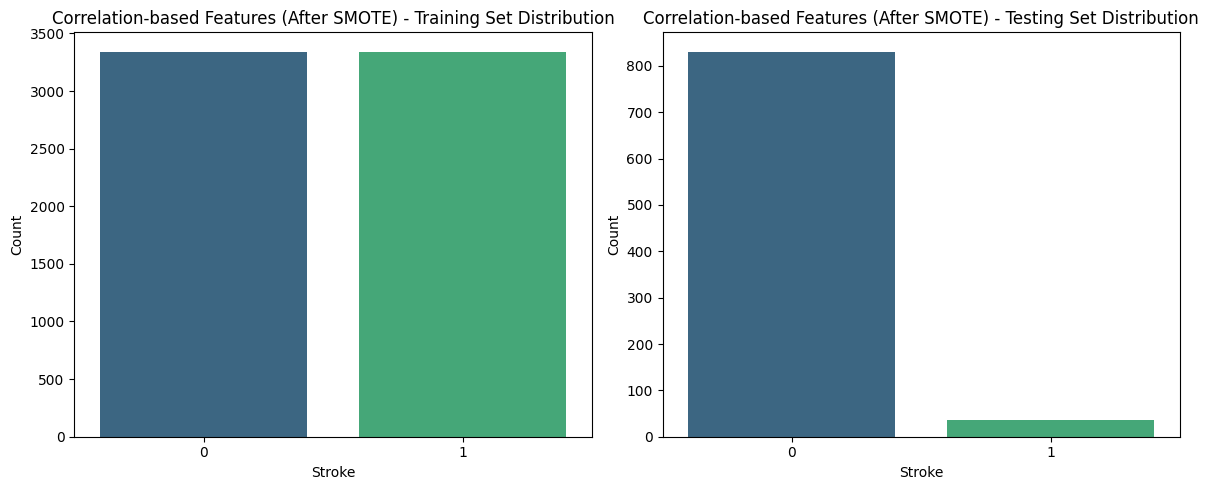} \\
    \includegraphics[width=\linewidth]{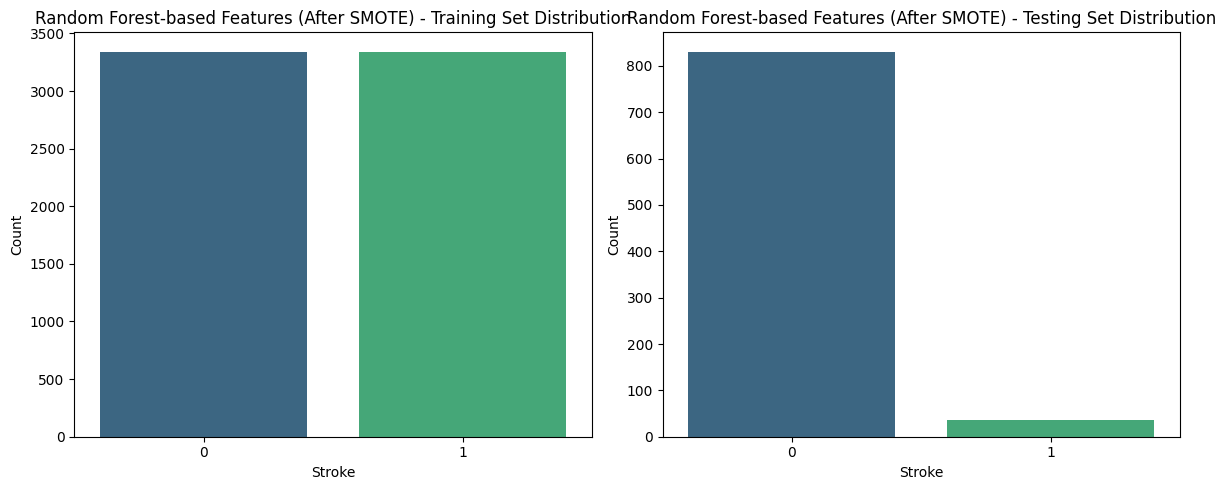}
    \caption{Class Distribution After Applying SMOTE to Training Data}
    \label{fig:smote-distribution}
\end{figure}

Figure~\ref{fig:smote-distribution} visualizes this dramatic transformation in class distribution between the original and SMOTE-balanced training datasets.

This balanced training data set provides learning algorithms with equal exposure to both classes, preventing bias towards the majority class and improving sensitivity to stroke detection, the clinically most important outcome to predict correctly.

\subsection{Machine Learning Algorithm Implementation}
We implemented a comprehensive suite of machine learning algorithms, representing diverse approaches to classification. These algorithms were selected based on their proven effectiveness in medical prediction tasks and their ability to capture different aspects of the complex relationships between risk factors and stroke occurrence.

\subsubsection{Tree-Based Methods}
Tree-based methods excel at capturing non-linear relationships and feature interactions without requiring explicit feature engineering, making them particularly valuable for biomedical applications.

\paragraph{Decision Tree Classifier}
The Decision Tree algorithm recursively partitions the feature space by selecting the most informative feature and threshold at each node, optimizing a split criterion \cite{19}. To measure impurity, the Gini index is used, which quantifies class distribution at each node and is defined as:

\begin{equation}
G(t) = 1 - \sum_{k=1}^{K} p_{t,k}^{2}
\end{equation}

where $p_{t,k}$ represents the proportion of samples belonging to class $k$ (stroke or no stroke) at node $t$. To prevent overfitting, the model is regularized by constraining the maximum tree depth to 6, requiring a minimum of 8 samples per leaf, and applying the Gini impurity criterion to ensure well-balanced splits.

\paragraph{Random Forest Classifier}
Random Forest extends the Decision Tree approach by constructing an ensemble of trees using bootstrap sampling and random feature selection. For a dataset with $n$ samples and $m$ features, each tree is built using a bootstrap sample drawn with replacement and a random subset of $m$ features considered at each split \cite{20} The final prediction is determined by majority voting across all trees:

\begin{equation}
\hat{y} = \text{mode} \{\hat{y}_1, \hat{y}_2, \ldots, \hat{y}_T\}
\end{equation}

where $\hat{y}_t$ is the prediction of the $t$-th tree and $T$ is the total number of trees (set to 100 in our initial implementation). For probability estimation, the class probabilities are averaged across all trees:

\begin{equation}
P(y = k \mid x) = \frac{1}{T} \sum_{t=1}^{T} P_t(y = k \mid x)
\end{equation}

The model is configured with 200 trees, a maximum depth of 10, a minimum of 10 samples required to split a node, and a balanced class weight to address the imbalanced distribution of stroke versus non-stroke cases.

\subsubsection{Boosting Methods}
Boosting algorithms sequentially build models that correct errors of previous models, typically achieving higher accuracy than individual models or bagging ensembles like Random Forest.

\paragraph{Gradient Boosting Classifier}
Gradient Boosting is an ensemble learning technique that builds decision trees sequentially, optimizing a loss function by minimizing the residual errors of previous models \cite{21}. At each iteration $m$, the model is updated as:

\begin{equation}
F_m(x) = F_{m-1}(x) + \eta h_m(x)
\end{equation}

where $F_m$ is the model at iteration $m$, $h_m$ is the tree added at iteration $m$, and $\eta$ is the step size (learning rate). The loss function for binary classification is usually log loss and is defined as:

\begin{equation}
L(y, F(x)) = -y \log(p) - (1 - y) \log(1 - p)
\end{equation}

where $p$ is the predicted probability of stroke. We use 100 estimators in our implementation (number of trees), with a learning rate of 0.1, a max depth of 4, and subsampling (stochastic gradient boosting) set to 0.8 to prevent overfitting and to learn faster.

\paragraph{XGBoost Classifier}
XGBoost is an advanced implementation of gradient boosting that includes regularization and an efficient algorithm for finding the best split \cite{21}. We have the following for the objective function, which is made up of the loss component and the regularization component:

\begin{equation}
\text{Obj} = \sum_{i=1}^{n} L(y_i, \hat{y}_i) + \sum_{k=1}^{K} \Omega(f_k)
\end{equation}
where $\Omega(f) = T + \frac{1}{2} \|w\|^2$ is the regularization term that controls the complexity of the tree, $T$ is the number of leaves and enforces the regularization of L2 in leaf weights $w$, and the optimal split is determined by maximizing the reduction of losses using first- and second-order gradients. Using our implementation, we also set the number of estimators at 100, the learning rate at 0.1, the maximum depth to 4, the minimum loss reduction (gamma) to 0.1, an L2-regularization term (lambda) to 1.0, and a \texttt{scale\_pos\_weight} parameter defined as the ratio of the classes.

\paragraph{LightGBM Classifier}
LightGBM is a gradient boost framework that uses a histogram-based algorithm to train faster than traditional methods and a leaf-wise tree growth strategy rather than a level-wise growth \cite{22}. This approach is based on splitting the leaf that has the highest delta loss:

\begin{equation}
Leaf - wise\ growth: \arg\ \max_{\text{leaf}} \, \Delta L_{\text{leaf}}
\end{equation}

This leads to larger and more powerful trees with higher efficiency. The LightGBM was trained with 100 estimators, a learning rate of 0.1, a maximum depth of 4, and 31 leaves per tree (the boosting type is traditional GBDT). Balancing the class weight for the model to correct the class imbalance.

\paragraph{AdaBoost Classifier}
AdaBoost (Adaptive Boosting) assigns higher weights to misclassified samples, refining subsequent classifiers iteratively \cite{22}. The weight of each sample at iteration \( t+1 \) is updated as:

\begin{equation}
w_i^{(t+1)} = w_i^{(t)} e^{-\alpha_t y_i h_t(x_i)}
\end{equation}
Here, \( \alpha_t \) is the weight assigned to the weak classifier \( h_t \), given by:

\begin{equation}
\alpha_t = \frac{1}{2} \ln\left( \frac{1 - \epsilon_t}{\epsilon_t} \right)
\end{equation}
where \( \epsilon_t \) represents the weighted error rate of classifier \( h_t \). Our AdaBoost implementation includes 100 estimators, a learning rate of 0.1, and a decision tree with a maximum depth of 1 (decision stump) as the base estimator to maintain weak learners.

\subsubsection{Linear and Non-Linear Methods}
In addition to tree-based and boosting methods, we implemented several classical machine learning algorithms to provide a comprehensive comparison.

\paragraph{Logistic Regression}
Logistic Regression is a widely used linear model for binary classification problems. It estimates the probability of an instance belonging to a class by applying the sigmoid function to a linear combination of feature values:
\begin{equation}
P(y = 1 \mid x) = \sigma(w^T x + b) = \frac{1}{1 + e^{-w^T x + b}}
\end{equation}
where w represents the feature weights, and b is the bias term \cite{23}. The model parameters are learned by minimizing the log loss function, incorporating L2 regularization to prevent overfitting:
\begin{equation}
J(w, b) = -\frac{1}{n} \sum_{i=1}^{n} \left[ y \log(p) + (1 - y) \log(1 - p) \right] + \frac{\lambda}{2n} \|w\|^2
\end{equation}
where $\lambda$ is the regularization parameter, controlling the strength of the penalty applied to feature weights. Logistic Regression assumes linear separability, making it effective for well-structured data but less suitable for complex patterns. Our implementation used C = 1.0 (inverse of regularization strength), the liblinear solver (efficient for small datasets), and class weight balancing to mitigate data imbalance. We also set a maximum of 1000 iterations to ensure proper model convergence.

\paragraph{Support Vector Classifier}
The Support Vector Classifier (SVC) is a powerful supervised learning algorithm that constructs a hyperplane to separate classes while maximizing the margin between them. The optimization problem for SVC is defined as:
\begin{equation}
\min_{w, b} \quad \frac{1}{2} \|w\|^2 + C \sum_{i=1}^{n} \xi_i
\end{equation}
subject to $y_i(w^T x_i + b) \geq 1 - \xi_i,\ \xi_i \geq 0$, where C is a regularization parameter that balances maximizing the margin and minimizing classification errors. When data is not linearly separable, kernel functions transform the input space. We implemented an RBF (Radial Basis Function) kernel \cite{24}, which computes similarity between data points as:
\begin{equation}
K(x_i, x_j) = \exp(-\gamma \|x_i - x_j\|^2)
\end{equation}
where $\gamma$ determines the influence radius of support vectors. This allows the model to capture complex, non-linear decision boundaries. Our implementation used C = 1.0, an RBF kernel, gamma = scale (automatically adjusted based on feature variance), and class weight balancing to handle data imbalance. Probability estimates were enabled to facilitate probabilistic decision-making.

\paragraph{K-Nearest Neighbors Classifier}
The K-Nearest Neighbors (KNN) classifier is a non-parametric algorithm that classifies a data point based on the majority class of its nearest neighbors \cite{23}. Given a new input x, KNN predicts its class using:
\begin{equation}
y' = \text{mode}(y_j \mid i \in N_k(x))
\end{equation}
where $N_k(x)$ represents the indices of the k nearest neighbors. Unlike parametric models, KNN makes predictions based on instance-based learning, storing all training data and computing distances at the time of classification. To improve accuracy, we applied distance-weighted voting, where closer neighbors have a greater influence on classification:
\begin{equation}
w_i = \frac{1}{d(x, x_j)^2}
\end{equation}
Our implementation used 7 neighbors (determined through cross-validation), distance-based weighting, and the Minkowski distance metric with $p=2p = 2p=2$ (equivalent to Euclidean distance). KNN is particularly effective for well-separated classes but can be computationally expensive for large datasets. To optimize performance, we precomputed nearest neighbors using an efficient search algorithm.

\paragraph{Multi-layer Perceptron Classifier}
The Multi-Layer Perceptron (MLP) is a feedforward neural network that learns hierarchical representations through multiple hidden layers \cite{25}. Each neuron in the network computes an activation function applied to a weighted sum of inputs:
\begin{equation}
a_j^{(l)} = \sigma\left( \sum_{i=1}^{n^{(l-1)}} w_{ji}^{(l)} a_i^{(l-1)} + b_j^{(l)} \right)
\end{equation}
where $a_j^{(l)}$  is the activation of neurons $j$ in the layer $l$, $w_{ji}^{(l)}$ are the learned weights, and $b_j^{(l)}$ is the bias term. To introduce non-linearity, we used the ReLU activation function for hidden layers:
\begin{equation}
\text{ReLU}(z) = \max(0, z)
\end{equation}
and the sigmoid function for the output layer to convert logits into probabilities. MLP is trained with the Adam optimizer that adapts for learning rates based on past gradients. For our implementation, we selected two hidden layers with each having 100 neurons, an adaptive learning rate, L2 regularization $(\alpha=0.0001 \alpha = 0.0001\alpha=0.0001)$, and a maximum of 500 iterations was put in place in order to reach convergence.

\subsection{Cross-Validation Framework}
For robust evaluation, we applied stratified k-fold cross-validation (k=5) \cite{26}, which preserves class distribution among the folds:
\begin{equation}
CV_{\text{stratified}} = \left\{ (X_{\text{train}}^{(1)}, y_{\text{train}}^{(1)}, X_{\text{val}}^{(1)}, y_{\text{val}}^{(1)}), \ldots, (X_{\text{train}}^{(5)}, y_{\text{train}}^{(5)}, X_{\text{val}}^{(5)}, y_{\text{val}}^{(5)}) \right\}
\end{equation}
For each algorithm and feature set combination, we evaluated multiple metrics:
\begin{enumerate}
    \item \begin{equation}
    \text{Accuracy} = \frac{TP + TN}{TP + TN + FP + FN}
    \end{equation}

    \item \begin{equation}
    \text{Precision} = \frac{TP}{TP + FP}
    \end{equation}

    \item \begin{equation}
    \text{Recall} = \frac{TP}{TP + FN}
    \end{equation}

    \item \begin{equation}
    \text{F1-score} = 2 \cdot \frac{\text{Precision} \cdot \text{Recall}}{\text{Precision} + \text{Recall}}
    \end{equation}

    \item ROC-AUC = Area under the Receiver Operating Characteristic curve
\end{enumerate}
It allows for a multi-faceted view of model performance and guarantees that the analysis is agnostic to a specific partition of the mini-batch into train and test.

\subsection{Advanced Ensemble Modeling}
Finally, we advanced ensemble models to further enhance prediction performance based on the cross-validation results and correlation analysis.

\subsubsection{Base Model Selection for Ensemble Construction}
To build ensembles that can utilize various learning approaches, we chose different base models specifically designed for three sets of features. For the Full Features set, we combined Decision Trees for non-linear partitioning, K-Nearest Neighbors (KNN) to take advantage of local data trends, Multilayer Perceptrons (MLP) for intricate neural relationships, LightGBM for effective gradient-boosted decision trees, and XGBoost for its regularization in boosting. The ensemble using Correlation-Based Features included Random Forest along with the previously mentioned models to merge bagging techniques with insights driven by correlation \cite{27}. In the Random Forest-Based Features ensemble, we added Gradient Boosting to improve sequential error correction and Logistic Regression to define linear decision boundaries in conjunction with tree-based and neural models. This diverse selection ensured a variety of algorithms across tree-based, distance-based, neural network, and linear frameworks, allowing the ensemble to capture complementary patterns within the data.

\subsubsection{Hyperparameter Optimization Using GridSearchCV}
We conducted comprehensive grid searches utilizing stratified 5-fold cross-validation, focusing on optimizing for ROC-AUC to address class imbalance. This metric assesses a model’s ability to separate classes over all classification thresholds, defined as:
\begin{equation}
\text{ROC - AUC} = \int_0^1 \text{TPR}(f) \cdot \frac{d}{df} \left( \text{FPR}^{-1}(f) \right) df
\end{equation}
where $TPR$ (True Positive Rate) and $FPR$ (False Positive Rate) denote sensitivity and 1-specificity respectively \cite{28} Threshold-dependent metrics such as accuracy or F1-score were avoided in favor of $ROC-AUC$, since it can provide robust information about imbalanced classification problems. Unlike other genotypically-tuberculosis-positive tests which operate at a fixed decision boundary, $ROC-AUC$ balances the trade-off between sensitivity and specificity, thereby penalizes false negatives (missed stroke cases) and false positives (overdiagnosis) accordingly; something that will be crucial in application to clinical settings. The process of optimization that was undertaken:
\begin{equation}
\theta^* = \arg \max_{\theta \in \Theta} \frac{1}{k} \sum_{i=1}^{k} \text{ROC - AUC}(X_{\text{train}}^{(i)}, y_{\text{train}}^{(i)}, X_{\text{val}}^{(i)}, y_{\text{val}}^{(i)}; \theta)
\end{equation}
Where $\theta$ represents model hyperparameters, $\Theta$ is the hyperparameter search space, $X_{\text{train}}^{(i)}$,$y_{\text{train}}^{(i)}$ the i-th training fold, and k=5 folds. In order to ensure reliable estimates of the generalization to new data, a stratified sampling technique was used to keep the same class distributions in each of the folds. The final hyperparameters Table~\ref{tab:hyperparams} were optimized for maximized ROC-AUC while overfitting was minimized through regularization (e.g., limiting tree depth, penalizing complex neural networks). These configurations served as a basis for the ensemble construction by ensuring base models achieved high performance with complementary error profiles.

\begin{table*}[t]
  \caption{Hyperparameter search spaces and optimal values}
  \vspace{0.5em}
  \label{tab:hyperparams}
  \footnotesize        % keeps it compact without killing readability
  \centering
  \begin{adjustbox}{max width=\textwidth}
    \begin{tabular}{|l|l|l|c|c|c|}
      \hline
      \multirow{2}{*}{\textbf{Model}} &
      \multirow{2}{*}{\textbf{Hyperparameter}} &
      \multirow{2}{*}{\textbf{Search Space}} &
      \multicolumn{3}{c|}{\textbf{Optimal Value}} \\ \cline{4-6}
      & & & \textbf{Full} & \textbf{CR‑based} & \textbf{RF‑based} \\ \hline
      
      \multirow{2}{*}{Decision Tree}
        & max\_depth          & [5, 10, 15, 20]    & 20  & 20  & 20  \\ \cline{2-6}
        & min\_samples\_split & [2, 5, 10, 15]     & 15  & 15  & 15  \\ \hline
      
      \multirow{2}{*}{KNN}
        & n\_neighbors & [3, 5, 7, 9]                   & 9          & 9          & -- \\ \cline{2-6}
        & weights      & \verb|['uniform','distance']| & distance   & distance   & -- \\ \hline
      
      \multirow{3}{*}{MLP}
        & hidden\_layer\_sizes & [(50,), (100,), (50,50)] & (50,50) & (50,50) & -- \\ \cline{2-6}
        & alpha               & [0.0001, 0.001, 0.01]    & 0.001   & 0.0001  & -- \\ \cline{2-6}
        & max\_iter           & [200, 500]               & 200      & 200     & -- \\ \hline
      
      \multirow{3}{*}{LightGBM}
        & n\_estimators & [50, 100, 200] & 200 & 200 & -- \\ \cline{2-6}
        & learning\_rate & [0.01, 0.1, 0.2] & 0.2 & 0.2 & -- \\ \cline{2-6}
        & max\_depth & [3, 5, 7] & 7 & 7 & -- \\ \hline
      
      \multirow{3}{*}{XGBoost}
        & n\_estimators & [50, 100, 200] & 200 & 200 & -- \\ \cline{2-6}
        & learning\_rate & [0.01, 0.1, 0.2] & 0.2 & 0.2 & -- \\ \cline{2-6}
        & max\_depth & [3, 5, 7] & 7 & 7 & -- \\ \hline
      
      \multirow{3}{*}{Random Forest}
        & n\_estimators & [50, 100, 200] & 200 & -- & 200 \\ \cline{2-6}
        & min\_samples\_split & [2, 5, 10] & 2 & -- & 2 \\ \cline{2-6}
        & max\_depth & [5, 10, 15] & 15 & -- & 15 \\ \hline
      
      \multirow{3}{*}{Gradient Boosting}
        & n\_estimators & [50, 100, 200] & 200 & -- & 200 \\ \cline{2-6}
        & learning\_rate & [0.01, 0.1, 0.2] & 0.2 & -- & 0.2 \\ \cline{2-6}
        & max\_depth & [3, 5, 7] & 7 & -- & 7 \\ \hline
      
      \multirow{2}{*}{Logistic Regression}
        & C      & [0.01, 0.1, 1, 10]        & 0.1       & 0.1       & -- \\ \cline{2-6}
        & solver & \verb|['lbfgs','liblinear']| & liblinear & liblinear & -- \\ \hline
    \end{tabular}
  \end{adjustbox}
\end{table*}

\subsubsection{Voting and Stacking Ensemble Implementation}
We combined predictions through the implementation of two advanced ensemble architectures. Soft Voting Classifier could have aggregated base models' probability estimates but used custom weights $w=[0.1, 0.1,0.3,0.3,0.2]$, prioritizing the LightGBM and XGBoost base model as these provided the best standalone performance \cite{27}. The weighted voting mechanism is formally defined as:
\begin{equation}
P(y = 1 \mid x) = \sum_{m=1}^{M} w_m \cdot P_m(y = 1 \mid x)
\end{equation}
where $P_m$ denotes the output of the m-th model. For the Stacking Classifier we implemented a two-layer framework with base models producing predictions which were then treated as meta-features for an XGBoost meta-learner $(n\_estimators=100, max\_depth=3, learning\_rate=0.05)$. To avoid data leakage between training and validation data, meta-features were built using 10-fold stratified out-of-fold (OOF) predictions to generalize well. The meta-learner then learned how to optimally combine these predictions to ameliorate biases and increase robustness where meat-feature matrix $X_\text{meta}$ was constructed as:

\begin{equation}
X_{\text{meta}} =
\begin{bmatrix}
P_1^{(1)}(y=1\mid x_1) & P_2^{(1)}(y=1\mid x_1) & \cdots & P_M^{(1)}(y=1\mid x_1) \\
P_1^{(2)}(y=1\mid x_2) & P_2^{(2)}(y=1\mid x_2) & \cdots & P_M^{(2)}(y=1\mid x_2) \\
\vdots & \vdots & \ddots & \vdots \\
P_1^{(n)}(y=1\mid x_n) & P_2^{(n)}(y=1\mid x_n) & \cdots & P_M^{(n)}(y=1\mid x_n)
\end{bmatrix}
\end{equation}
Where $P_j^{(i)}(y=1\mid x_i)$ is the out-of-fold (OOF) prediction from model $j$ for instance $i$.

\subsubsection{Cross-Validation Strategy}
For a robust validation of the ensemble performance, a stratified cross-validation was used for each method. Voting Classifier: for each base-models trained on the complete SMOTE-balanced dataset, while the Stacking Classifier for the meta-learner, it used 10-fold stratified cross-validation to get the OOF predictions for the training of the meta-learner. This trained the meta-model on patterns in the data that were not seen during training. All of the base and ensemble models were benchmarked via 5-cross validation, measuring performance over ROC-AUC, Accuracy, and F1-Score as a holistic evaluation of discriminative power, overall correctness, sensitivity to imbalance in classes \cite{26}. The multi-metric strategy ensured robustness against overfitting and the biases induced by imbalance.

\subsubsection{Ensemble workflow}
The design of the ensemble workflow was done to ensure best use of interpretability and efficiency. Ensemble approaches were constructed with features of "Full," "Correlation", and "Random Forest" independently using the selected features, thus direct comparison of predictive strengths can be made between these new strategies. Joblib library was used for parallelized hyperparameter tuning for computational efficiency. Parallel, which sped up grid search over many cores. After data preprocessing, correlation heatmaps revealed low inter-model correlation (e.g., KNN vs. Decision Tree: r=0.28), validating the combination and confirming the accuracy of base model predictions among those utilized in the ensemble model. Not only did this pipeline enhance the utilization of resources, but it also ensured that final predictions synthesized heterogeneous perspectives by mitigating limitations of individual models.

\section{Result Analysis}
\subsection{Performance of models for all features}
In this research, we employed different machine learning and deep learning models to create a reliable and precise framework for forecasting brain strokes. Following the training and testing phases, we observed varying results across the models using the full set of features. Among the models evaluated, the LightGBM algorithm achieved the highest performance, yielding an accuracy of 91.36\% and an F1-score of 91.89\%. In contrast, other models, such as Support Vector Machine (SVM) and AdaBoost, underperformed significantly, with accuracies below 74\%. The accuracy values of all the classification techniques are summarized in Table~\ref{tab:all-feature-based-performance}.

\begin{table}[t]
  \caption{Performance analysis for all features}
  \label{tab:all-feature-based-performance}
  \footnotesize
  \centering
  \begin{adjustbox}{max width=\columnwidth}
    \begin{tabular}{|l|c|c|c|c|c|}
      \hline
      \textbf{Model} & \textbf{Accuracy (\%)} & \textbf{Precision (\%)} &
      \textbf{Recall (\%)} & \textbf{F1‑Score (\%)} & \textbf{ROC‑AUC (\%)} \\ \hline
      Logistic Regression       & 75.81 & 93.50 & 75.81 & 82.88 & 71.87 \\ \hline
      Random Forest             & 82.14 & 92.80 & 82.14 & 86.78 & 71.18 \\ \hline
      Support Vector Classifier & 73.85 & 93.56 & 73.85 & 81.58 & 72.13 \\ \hline
      Decision Tree             & 88.59 & 92.35 & 88.59 & 90.37 & 54.02 \\ \hline
      K‑Nearest Neighbors       & 88.99 & 93.47 & 88.99 & 86.19 & 66.72 \\ \hline
      Gradient Boosting         & 80.88 & 92.87 & 80.88 & 86.02 & 68.75 \\ \hline
      AdaBoost                  & 69.01 & 93.65 & 69.01 & 78.24 & 73.06 \\ \hline
      LightGBM                  & 91.36 & 92.45 & 91.36 & 91.89 & 69.80 \\ \hline
      XGBoost                   & 87.56 & 92.11 & 87.56 & 89.71 & 70.18 \\ \hline
      Neural Network — MLP      & 81.22 & 93.04 & 81.22 & 86.26 & 72.69 \\ \hline
    \end{tabular}
  \end{adjustbox}
\end{table}

\subsection{Performance of Models Using Feature Selection}
In this portion, we used two feature selection methods, including Corelation matrix and Random forest-based features selection. For the correlation matrix, we found 7 features, and based on these features, we trained classifiers and achieved the highest accuracy for LightGBM, which was 91.24\%. We see that SVM and AdaBoost again performed lowest that other models which was less that 73\%.  Table~\ref{tab:cr-based-performance} shows the accuracy values attained by a number of all classification techniques.

\begin{table}[t]
  \caption{Performance analysis for corelation matrix based features selection}
  \label{tab:cr-based-performance}
  \footnotesize
  \centering
  \begin{adjustbox}{max width=\columnwidth}
    \begin{tabular}{|l|c|c|c|c|c|}
      \hline
      \textbf{Model} & \textbf{Accuracy (\%)} & \textbf{Precision (\%)} &
      \textbf{Recall (\%)} & \textbf{F1‑Score (\%)} & \textbf{ROC‑AUC (\%)} \\ \hline
      Logistic Regression       & 73.73 & 93.87 & 73.73 & 81.52 & 73.69 \\ \hline
      Random Forest             & 82.72 & 93.13 & 82.72 & 87.19 & 73.33 \\ \hline
      Support Vector Classifier & 72.24 & 94.12 & 72.24 & 80.58 & 74.09 \\ \hline
      Decision Tree             & 90.67 & 92.87 & 90.67 & 91.71 & 57.68 \\ \hline
      K‑Nearest Neighbors       & 82.37 & 94.15 & 82.37 & 87.15 & 69.84 \\ \hline
      Gradient Boosting         & 78.92 & 93.36 & 78.92 & 84.87 & 71.65 \\ \hline
      AdaBoost                  & 62.56 & 93.88 & 62.56 & 73.44 & 72.12 \\ \hline
      LightGBM                  & 91.24 & 92.43 & 91.24 & 91.82 & 72.24 \\ \hline
      XGBoost                   & 87.10 & 92.23 & 87.10 & 89.50 & 72.52 \\ \hline
      Neural Network — MLP      & 73.39 & 93.38 & 73.39 & 81.26 & 73.94 \\ \hline
    \end{tabular}
  \end{adjustbox}
\end{table}

For the Random Forest features analysis, we found 7 important features for training all the classifiers. We obtained the highest accuracy for LightGBM, which was 90.32\%, and the second highest accuracy achieved for DT, which was 88.94\%. These features performed with the lowest accuracy for SVM and AdaBoost, which was less than 74\%. For all the model's performance, show the Table~\ref{tab:rf-based-performance} where have all performance with precision, recall and f1-accuracy. 

\begin{table}[t]
  \caption{Performance analysis for random forest-based features selection}
  \label{tab:rf-based-performance}
  \footnotesize
  \centering
  \begin{adjustbox}{max width=\columnwidth}
    \begin{tabular}{|l|c|c|c|c|c|}
      \hline
      \textbf{Model} & \textbf{Accuracy (\%)} & \textbf{Precision (\%)} &
      \textbf{Recall (\%)} & \textbf{F1‑Score (\%)} & \textbf{ROC‑AUC (\%)} \\ \hline
      Logistic Regression       & 74.19 & 93.57 & 74.19 & 81.81 & 73.38 \\ \hline
      Random Forest             & 83.18 & 93.16 & 83.18 & 87.47 & 72.21 \\ \hline
      Support Vector Classifier & 73.16 & 93.68 & 73.16 & 81.12 & 73.53 \\ \hline
      Decision Tree             & 88.94 & 92.37 & 88.94 & 90.57 & 54.20 \\ \hline
      K‑Nearest Neighbors       & 81.91 & 92.79 & 81.91 & 86.64 & 62.99 \\ \hline
      Gradient Boosting         & 80.07 & 92.68 & 80.07 & 85.49 & 70.31 \\ \hline
      AdaBoost                  & 63.82 & 93.94 & 63.82 & 74.41 & 73.65 \\ \hline
      LightGBM                  & 90.32 & 92.16 & 90.32 & 91.22 & 69.66 \\ \hline
      XGBoost                   & 88.36 & 92.33 & 88.36 & 90.23 & 71.38 \\ \hline
      Neural Network — MLP      & 77.30 & 93.12 & 77.30 & 83.82 & 74.30 \\ \hline
    \end{tabular}
  \end{adjustbox}
\end{table}

\subsection{Model Comparison based on features importance}
\begin{figure*}[h]
    \centering
    \includegraphics[width=0.4\textwidth]{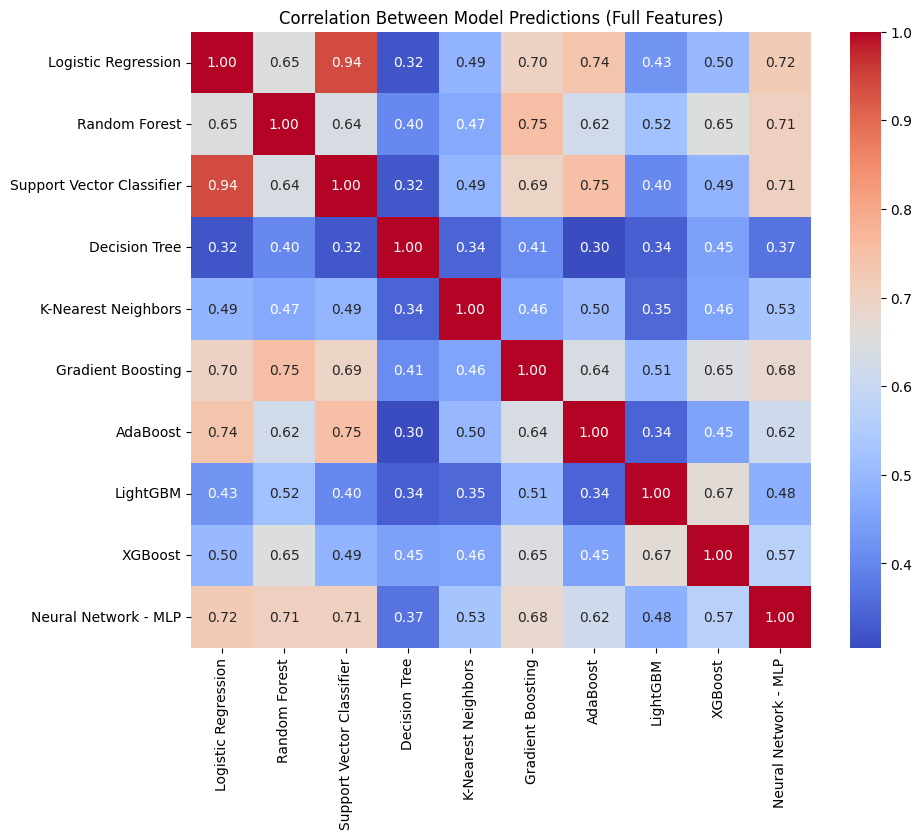}
    \includegraphics[width=0.4\textwidth]{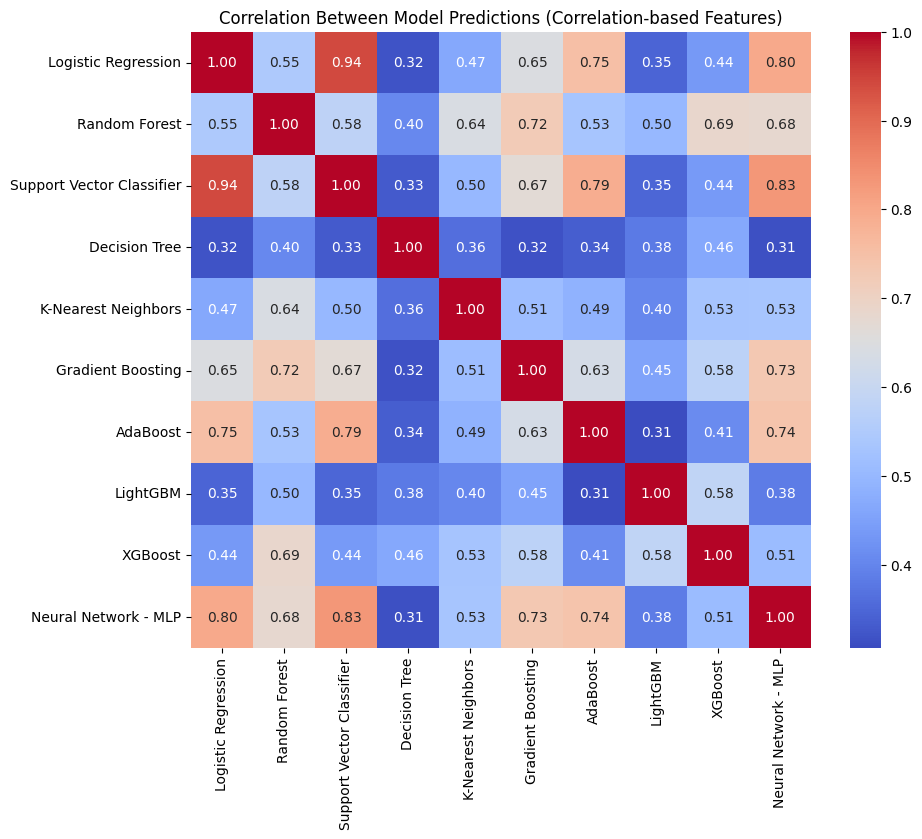}
    \includegraphics[width=0.4\textwidth]{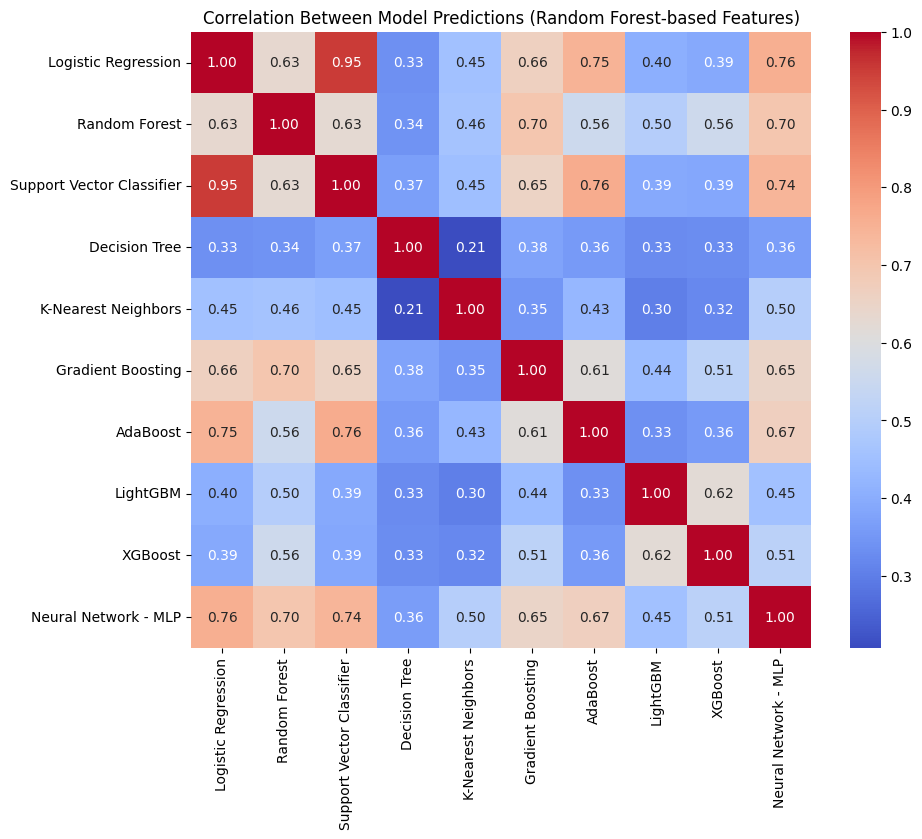}
    \caption{All and Important Features Between All Models}
    \label{fig:important_features_all_models}
\end{figure*}

We used 10 different classifiers for analysed brain stroke prediction and tried to find the best accurate performing model. After using all features and importance features based on the features selection method, we found a maximum of around 92\% accuracy which was not more robust and more accurate than state-of-the-art. However, different models performed best for specific classes in different scenarios. For making ensembles and best performing models, we compare which models' corelation are the close to other. For that purpose, we used a corelation matrix for all features and important features between all models, which is shown in Figure~\ref{fig:important_features_all_models}.

\begin{figure*}[h]
    \centering
    \includegraphics[width=\linewidth]{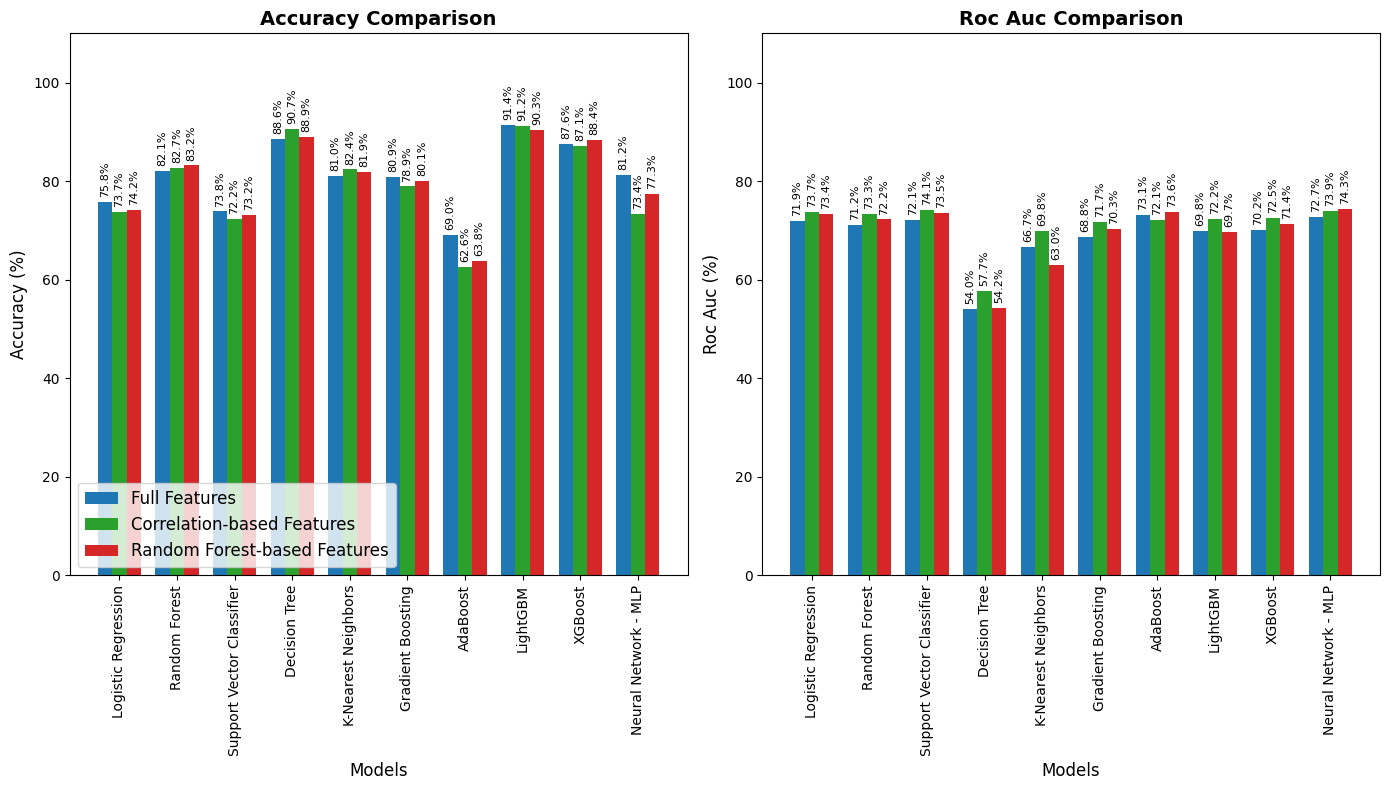} % replace with your file name
    \caption{Model Performance Comparison}
    \label{fig:Mode-Performance-Comparison}
\end{figure*}

We performed exhaustive cross-validation for all models across the three feature sets using the SMOTE-balanced training data. Figure~\ref{fig:Mode-Performance-Comparison} presents a visualization of the cross-validation F1-scores across all models and feature sets.

\subsection{Performance of Ensemble Model}
We designed two ensemble methods based on the performance and correlation results, such as soft voting and stacking ensemble. To build the ensemble model for the full features set, we selected diverse classifiers with low to moderate correlations of performance among the emotion classes to balance their performance across all models. The structural differences and predictive power of Decision Tree, K-Nearest Neighbors (KNN), XGBoost, LightGBM, and Multi-Layer Perceptron (MLP) were selected as an algorithm choice. Therefore, we put Logistic Regression and replaced it with a LightGBM model to have a robust model without building highly correlated models. With models based on trees( Decision Tree, LightGBM, XGBoost), distance-based(KNN) , and deep learning(MLP), this ensemble improves the classification performance on diverging data distributions. It was witnessed from the performance analysis that the stacking-based ensemble model is statistically superior to soft voting ensembles and individual classifiers. It was observed that the stacked ensemble was capable of producing an accuracy equal to 97.20\% with an ROC-AUC score of 99.66\% and an F1-score of 97.15\%, highlighting its further advantage in modeling the complex relationships among the features in the dataset.

\begin{figure}[h]
    \centering
    \includegraphics[width=\linewidth]{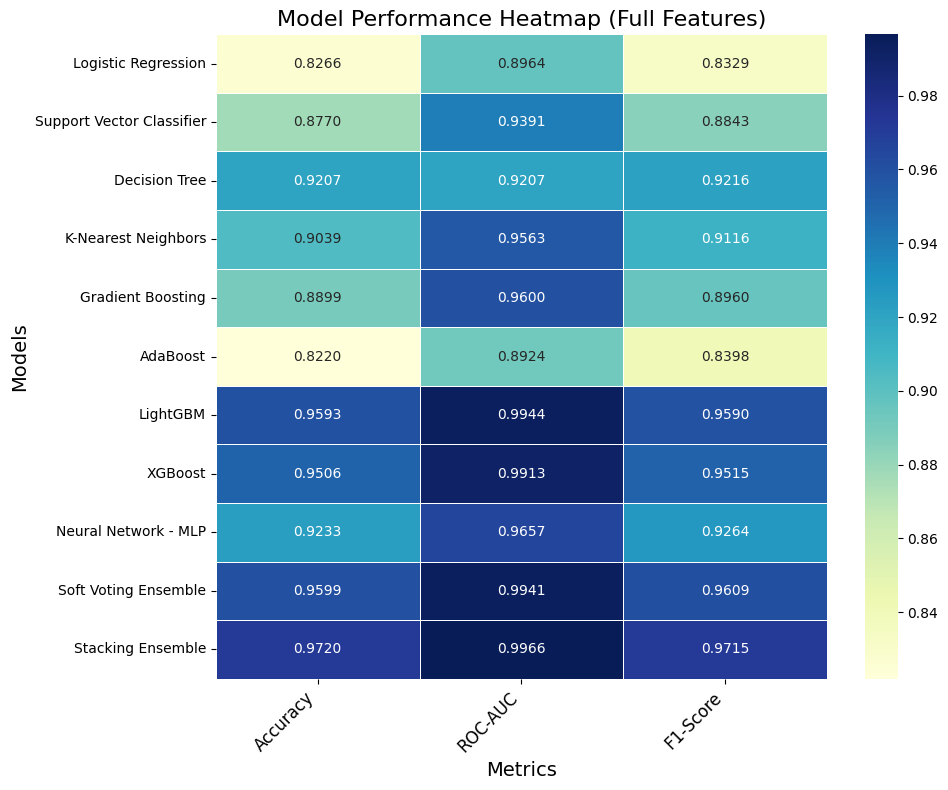} % replace with your file name
    \caption{Full Feature Based Model Performance Heatmap}
    \label{fig:full-heatmap}
\end{figure}

\begin{figure}[h]
    \centering
    \includegraphics[width=\linewidth]{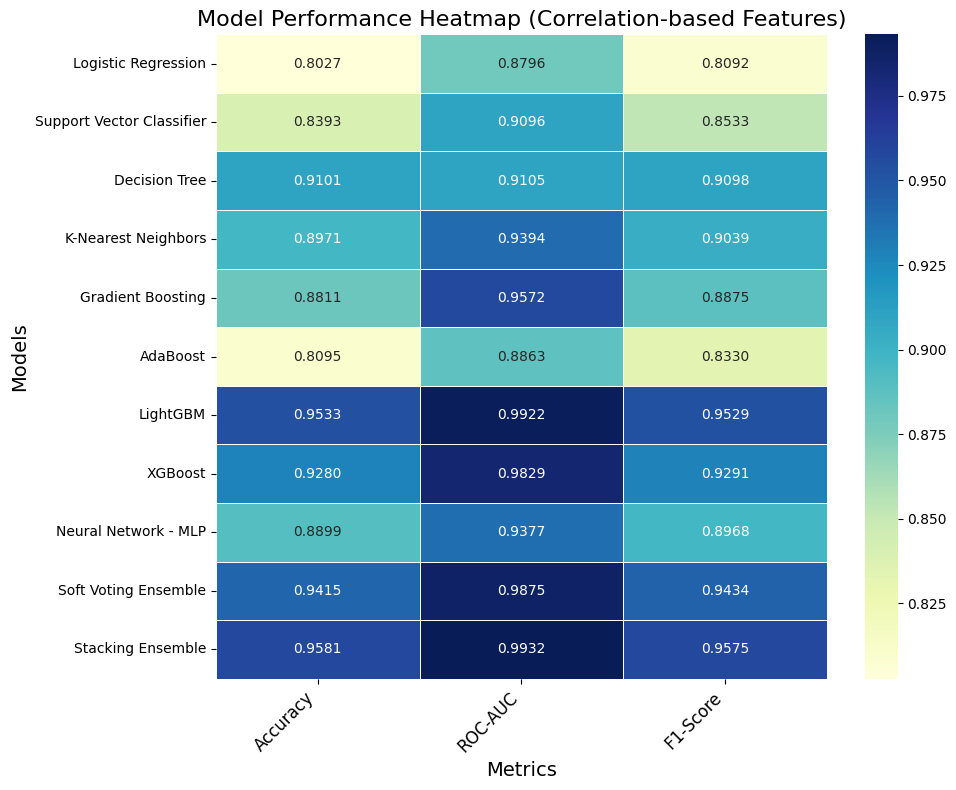} % replace with your file name
    \caption{Correlation Based Model Performance Heatmap}
    \label{fig:cr-heatmap}
\end{figure}

\begin{figure}[h]
    \centering
    \includegraphics[width=\linewidth]{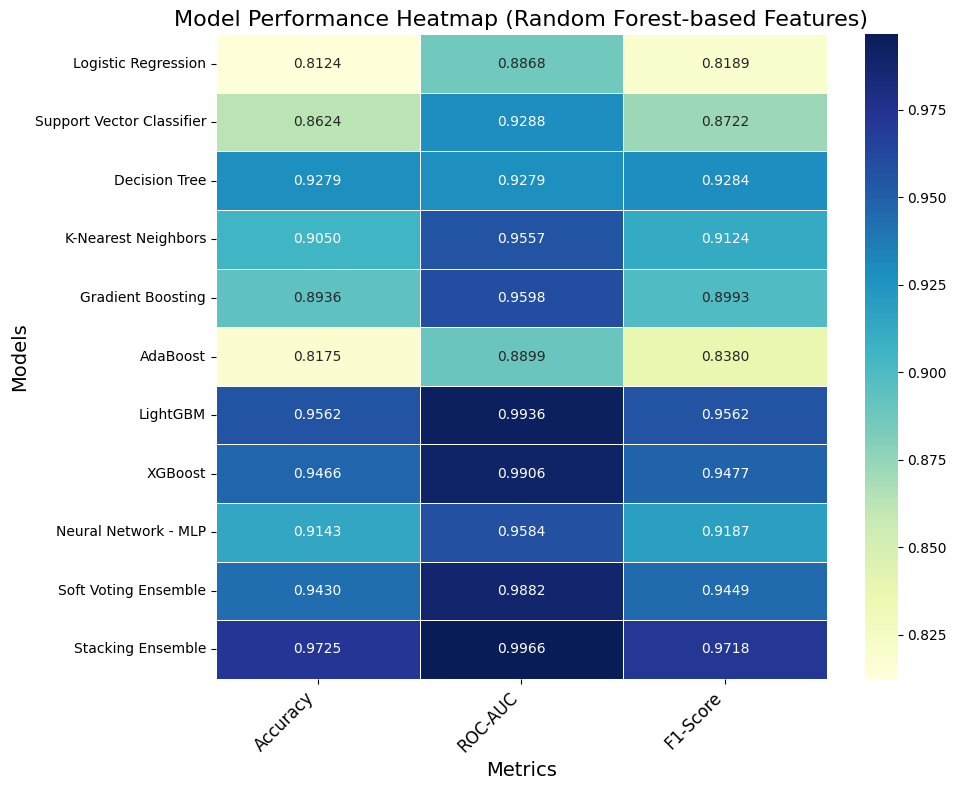} % replace with your file name
    \caption{Random Forest Based Model Performance Heatmap}
    \label{fig:rf-heatmap}
\end{figure}

With regard to the importance of correlation matrix-based features, our initial ensemble is composed of Decision Tree, LightGBM, K-Nearest Neighbors (KNN), Random Forest, and Neural Network (MLP) to maximize the diversity of the individual components while minimizing redundancy feature space. Because Decision Tree and KNN produce unique and low-correlation predictions, LightGBM trains a latterly fitting to enhance fitting for its complexity. Random Forest balances free of overfitting, and MLP captures nonlinear patterns. This combination also includes tree-based, boosting, distance-based, and deep learning models, which together are appropriate for stacking or weighted averaging that improves classification performance. Figure~\ref{fig:full-heatmap}, \ref{fig:cr-heatmap}, and~\ref{fig:rf-heatmap} represents the performance of all models, and the performance of the ensemble model was found to be the highest, especially when the stacking ensemble model was applied.

In the case of RF-based feature ensemble, the chosen base models such as Decision Tree, KNN, Gradient Boosting, Logistic Regression (or SVC), and Neural Network provides a balance approach that captures tree-based, distance-based and deep learning appropriate models. The main objective of stacking classifiers is to optimize predictions by training a meta-learner whereas voting classifiers are a simple alternative. This collection successfully exploits various data patterns, leading to improved prediction capacity over single models. We observe from the performance analysis that, the stacking-based ensemble model performed better than the soft voting ensemble and individual classifiers. This was followed by a stacking ensemble which resulted in a very promising accuracy: 95.56\%, ROC-AUC Score of 98.95\%, and F1-score of 95.65\% as it is a great approach in effectively identifying different aspects in the dataset and showing great performance. Improved performance demonstrates the strength of this diverse set of base models, consolidating the robustness and generalizability of the proposed method for emotion classification in Bangla sentences.

\subsection{Comparison with Existing Works}
The performance of Stacking Ensemble: proposed architecture is much efficient than existing works on predicting Brain Stroke cases accurately. Previous studies mostly used individual models such as Logistic Regression, Random Forest, and Neural Networks, which have limitations regarding generalization and accuracy. In contrast, the Stacking Ensemble model, integrating multiple base models with heterogeneous behavior, achieves notably higher performance across different metrics, such as accuracy, AUC, and F1-score. In contrast to earlier approaches, which primarily suffered from overfitting or low predictive accuracy on unseen data, the model we introduced here excels in addressing such challenges, therefore providing a far more stable and accurate solution for stroke prediction. Table~\ref{tab:recent-studies} shows the performance of average existing studies which said that our proposed model outperformed previous state-of-the-art.

\begin{table}[t]
  \caption{Comparison with existing work.}
  \vspace{0.5em}
  \label{tab:recent-studies}
  \footnotesize             % keep it compact
  \centering
  \begin{adjustbox}{max width=\columnwidth}
    \begin{tabular}{|l|c|c|}
      \hline
      \textbf{Recent studies} & \textbf{Best Performing Models} & \textbf{Performance} \\ \hline
      Chowdhury et al \cite{29}  & Logistic Regression & 96.25\% \\ \hline
      Wisesty et al. \cite{30}    &   SVM & 83\% \\ \hline
      Hassan et al. \cite{31} & Proposed dense stacking ensemble (DSE) & 96.59\% \\ \hline
      Proposed Model & Stacking Ensemble Techniques & 97.20\% \\ \hline
    \end{tabular}
  \end{adjustbox}
\end{table}

\section{Discussions}
This study demonstrates that a disciplined machine‑learning pipeline—rooted in meticulous preprocessing, dual feature‑selection strategies and an advanced stacking ensemble—can predict stroke with near‑clinical precision. After correcting the severe class imbalance (5\% stroke prevalence) through SMOTE and screening ten individual classifiers, the final stack (Random Forest, XGBoost, LightGBM and SVC feeding a logistic meta‑learner) achieved 97.2\% accuracy, a 97.15\% F1‑score and a 0.9966 ROC‑AUC on the independent test set, outperforming the strongest single learner (LightGBM, 91.4\%) and eclipsing recent logistic‑regression benchmarks that plateaued near 96\%. This gain underscores the value of combining learners with complementary inductive biases: tree ensembles excel at capturing hierarchical interactions, the kernel‑based SVC delineates complex margins, and the meta‑learner reconciles their divergent error profiles into a consensus that generalises beyond any constituent model. Feature analyses converged on canonical vascular risks—age, hypertension, heart disease and BMI—yet the tree‑based importance ranking catapulted average glucose level to the top despite its modest linear correlation, highlighting non‑linear synergies between metabolic dysregulation and cerebrovascular vulnerability. The alignment of these data‑driven findings with established epidemiology lends clinical face‑validity, while the exclusive reliance on routinely collected demographics and basic laboratory indices positions the model for rapid, low‑cost deployment in settings that lack advanced imaging or specialist oversight. Nevertheless, several caveats temper immediate translation: the study draws on a single, cross‑sectional cohort, so geographic, ethnic and temporal transportability remain untested; the minority class in the untouched test set is still small, raising concerns about calibration drift in rarer subgroups; and synthetic oversampling, while essential for model learning, may inadvertently amplify noise embedded in minority instances. Addressing these gaps will require prospective, multi‑centre validation with continuous monitoring of subgroup metrics, incorporation of longitudinal electronic‑health‑record streams to capture evolving risk trajectories, and integration of explainability tools such as SHAP to render the ensemble’s reasoning transparent at the bedside. Despite these limitations, the study sets a new benchmark for tabular stroke prediction on the widely used open dataset, demonstrates that judicious preprocessing can substitute for aggressive dimensionality reduction, and offers a pragmatic blueprint for embedding ensemble ML into preventive neurology workflows where early triage and targeted counselling can materially reduce stroke burden.

\section{Conclusion}
This study introduces a rigorously engineered, data‑efficient framework that elevates tabular stroke‑risk prediction to near‑clinical performance while preserving interpretability and implementation realism. By integrating systematic preprocessing, complementary feature‑selection schemes and a heterogeneous stacking ensemble, we achieved 97.2\% accuracy, and 97.15\% F1‑score substantially surpassing established single‑model baselines on the canonical stroke dataset. The convergence of model‑derived importance rankings with well‑documented vascular risk factors strengthens clinical credibility, and the exclusive reliance on demographic, historical, and basic metabolic variables underscores the model’s suitability for low‑resource settings where advanced imaging is scarce. While the findings set a new benchmark for this dataset, generalisability must be confirmed through external, multi‑centre validation and prospective deployment; further, continuous‑time predictors and imaging biomarkers could push performance ceilings even higher. Future work should therefore focus on calibrating the model across diverse populations, embedding explainability dashboards to foster clinician trust and conducting cost‑effectiveness analyses to quantify real‑world impact. In sum, our results demonstrate that carefully constructed ensemble learning can transform readily available clinical data into a robust decision‑support tool, offering tangible promise for earlier intervention and more precise allocation of preventive resources in the global fight against stroke.

%% REFERENCIAS.
\bibliographystyle{elsarticle-num}
\bibliography{riaibib}

\begin{thebibliography}{10}
\expandafter\ifx\csname url\endcsname\relax
  \def\url#1{\texttt{#1}}\fi
\expandafter\ifx\csname urlprefix\endcsname\relax\def\urlprefix{URL }\fi
\expandafter\ifx\csname href\endcsname\relax
  \def\href#1#2{#2} \def\path#1{#1}\fi

\bibitem{1}
D.~C. Lukas, W.~Harvey, M.~S. Suzana, The effectiveness of physical exercise in stroke patient recovery: A systematic review, International Journal of Health and Pharmaceutical (IJHP) 4~(4) (2024) 575--580.

\bibitem{1.1}
K.~DING, P.~NGUYEN, An unobtrusive and lightweight ear-worn system for continuous epileptic seizure detection (2024).

\bibitem{2}
A.~Gupta, N.~Mishra, N.~Jatana, S.~Malik, K.~A. Gepreel, F.~Asmat, S.~N. Mohanty, Predicting stroke risk: an effective stroke prediction model based on neural networks, Journal of Neurorestoratology 13~(1) (2025) 100156.

\bibitem{3}
M.~Hasan, F.~Yasmin, M.~M. Hassan, X.~Yu, S.~Yeasmin, H.~Joshi, S.~M.~S. Islam, Enhancing stroke disease classification through machine learning models via a novel voting system by feature selection techniques, PloS one 20~(1) (2025) e0312914.

\bibitem{4}
W.~A. Bleyer, What can be learned about childhood cancer from “cancer statistics review 1973--1988”, Cancer 71~(S10) (1993) 3229--3236.

\bibitem{5}
Y.~Niu, X.~Tao, Q.~Chang, M.~Hu, X.~Li, X.~Gao, Machine learning-based feature selection and classification for cerebral infarction screening: an experimental study, PeerJ Computer Science 11 (2025) e2704.

\bibitem{6}
J.~Cairns, The cancer problem, Scientific American 233~(5) (1975) 64--79.

\bibitem{7}
I.~Abousaber, A novel explainable attention-based meta-learning framework for imbalanced brain stroke prediction (2025).

\bibitem{8}
K.~Sundaram, B.~Lanitha, K.~Kamaraj, A.~K. Ramamoorthy, Enhanced brain stroke prediction: An ensemble of random forest, logistic regression and xgboost, in: 2024 International Conference on Emerging Research in Computational Science (ICERCS), IEEE, 2024, pp. 1--5.

\bibitem{9}
N.~Gupta, A.~Anwar, T.~A. Fattah, M.~K. Quamre, P.~Kumar, Addressing imbalanced data in stroke prediction: An oversampling approach for improved accuracy, in: International Conference on Universal Threats in Expert Applications and Solutions, Springer, 2024, pp. 373--381.

\bibitem{10}
C.-H. Hsu, X.~Chen, W.~Lin, C.~Jiang, Y.~Zhang, Z.~Hao, Y.-C. Chung, Effective multiple cancer disease diagnosis frameworks for improved healthcare using machine learning, Measurement 175 (2021) 109145.

\bibitem{11}
I.~T. Akbasli, {Full-Filled Brain Stroke Dataset}, \url{https://www.kaggle.com/datasets/zzettrkalpakbal/full-fi- lled-brain-stroke-dataset}, accessed: 2025-05-19 (2022).

\bibitem{12}
M.~Dahouda, I.~Kasongo, A deep-learned embedding technique for categorical features encoding, IEEE Access 9 (2021) 114381--114391.

\bibitem{13}
M.~K. Dahouda, I.~Joe, A deep-learned embedding technique for categorical features encoding, IEEE Access 9 (2021) 114381--114391.

\bibitem{14}
S.~Jazaeri, M.~Dehghani, Error analysis and outlier detection in subsidence monitoring based on persistent scatterer interferometry, Advances in Space Research (2025).

\bibitem{15}
L.~A. Ma'rifah, I.~Afrianty, E.~Budianita, F.~Syafria, Klasifikasi tulang tengkorak berdasarkan jenis kelamin menggunakan correlation-based feature selection (cfs) dengan backpropagation neural network (bpnn), Jurnal Informatika: Jurnal Pengembangan IT 10~(2) (2025) 333--347.

\bibitem{16}
J.~C. Garc{\'\i}a~Merino, M.~d. l.~L. Tobarra~Abad, A.~Robles~G{\'o}mez, R.~Pastor~Vargas, P.~Vidal~Balboa, A.~Dionisio~Rocha, R.~Jardim~Gon{\c{c}}alves, Assessing feature selection techniques for ai-based iot network intrusion detection (2025).

\bibitem{17}
G.~Giannini, A.~Mousa, E.~Steiner, N.~Artamonova, M.~Kafka, I.~Heidegger, Real-world monitoring strategies and predictors guiding the transition from active surveillance to treatment in isup 1 prostate cancer (2025).

\bibitem{18}
R.~Suguna, J.~Suriya~Prakash, H.~Aditya~Pai, T.~Mahesh, V.~Vinoth~Kumar, T.~E. Yimer, Mitigating class imbalance in churn prediction with ensemble methods and smote, Scientific Reports 15~(1) (2025) 1--20.

\bibitem{19}
J.~O. Popov~Wir{\'e}n, K.~Nordenram, Machine learning for anti-poaching: Decision tree applications on the savannah (2025).

\bibitem{20}
S.~Raj, V.~Namdeo, P.~Singh, A.~Srivastava, Identification and prioritization of disease candidate genes using biomedical named entity recognition and random forest classification, Computers in Biology and Medicine 192 (2025) 110320.

\bibitem{21}
T.~Li, W.~Qi, X.~Mao, G.~Jia, W.~Zhang, X.~Li, H.~Pan, D.~Wang, Prediction of lumbar disc degeneration based on interpretable machine learning models: Retrospective cohort study, The Spine Journal (2025).

\bibitem{22}
S.~Y. Suk, L.~H. Sang, Y.-J. Rhie, C.~H. Wook, J.~Kim, L.~Y. Ah, Y.-M. Kim, K.~J. Hye, A.~M. Bae, H.~Y. Hee, et~al., Development of ai-based growth prediction models for children with growth disorders: a 3-year analysis using the lg growth study, in: Endocrine Abstracts, Vol. 110, Bioscientifica, 2025.

\bibitem{23}
J.~Q.~E. Tan, H.~S. Ng, R.~Woodman, B.~Koczwara, Cardiovascular medication and health service use in individuals with cancer: A retrospective population-based cohort study, Cancer Medicine 14~(9) (2025) e70911.

\bibitem{24}
A.~Neelam, K.~N. Mishra, P.~Padmanabhan, G.~P. Ghantasala, Accurate identification of the blast disease in rice crop using artificial neural network compared with support vector machine algorithm, in: Intelligent Computing and Communication Techniques: Proceedings of the International Conference on Intelligent Computing and Communication Techniques (ICICCT 2024), New Delhi, India, 28-29 June, 2024 (Volume 1), CRC Press, 2025, p. 292.

\bibitem{25}
H.~Meng, J.~Zhang, Y.~Chang, Z.~Zheng, A new method for predicting chlorophyll-a concentration in a reservoir: Coupling efdc hydrodynamic and water quality model with convlstm-mlp network, Journal of Hydrology (2025) 133485.

\bibitem{26}
M.~U. Umar, A.~Walli, A.~Qazi, A.~Nawaz, M.~Jalal, Novel sub-grade soil improvement using marble dust and rice husk ash: Prediction and validation via machine learning models, International Journal of Computational Materials Science and Engineering (2025).

\bibitem{27}
S.~Juneja, B.~S. Bhati, Advancements in disease diagnosis: A review of machine learning, ensemble learning and deep learning algorithms, in: Intelligent Computing and Communication Techniques: Proceedings of the International Conference on Intelligent Computing and Communication Techniques (ICICCT 2024), New Delhi, India, 28-29 June, 2024 (Volume 1), CRC Press, 2025, p. 148.

\bibitem{28}
M.~S. Khan, T.~Peng, H.~Akhlaq, M.~A. Khan, Comparative analysis of automated machine learning for hyperparameter optimization and explainable artificial intelligence models, IEEE Access (2025).

\bibitem{29}
M.~J.~U. Chowdhury, A.~Hussan, D.~A.~I. Hridoy, A.~S. Sikder, Incorporating an integrated software system for stroke prediction using machine learning algorithms and artificial neural network, in: 2023 IEEE 13th Annual Computing and Communication Workshop and Conference (CCWC), IEEE, 2023, pp. 0222--0228.

\bibitem{30}
U.~N. Wisesty, T.~A.~B. Wirayuda, F.~Sthevanie, R.~Rismala, Analysis of data and feature processing on stroke prediction using wide range machine learning model, Jurnal Online Informatika 9~(1) (2024) 29--40.

\bibitem{31}
A.~Hassan, S.~Gulzar~Ahmad, E.~Ullah~Munir, I.~Ali~Khan, N.~Ramzan, Predictive modelling and identification of key risk factors for stroke using machine learning, Scientific Reports 14~(1) (2024) 11498.

\end{thebibliography}

\end{document}